\documentclass[12pt]{article}
\usepackage[dvips]{graphicx}
\usepackage{psfrag}
\usepackage{pdproc}

  \makeatletter 
  \def\@cite#1{[#1]} 
  \makeatother    
\begin{document}

\renewcommand{\thefootnote}{\alph{footnote}}
\newcommand{\GeV}{\rm GeV}
\newcommand{\Rp}{\mbox{$\not \hspace{-0.15cm} R_p$}}
\title{
 Searches for SUSY and Exotics at HERA
}

\author{ 
JOHANNES HALLER
\footnote{now at CERN/PH-ATR, 1211 Geneva, Switzerland}
\footnote{on behalf of the H1 and ZEUS collaborations}
}

\address{Physikalisches Institut, Universit\"at Heidelberg 
 \\ {\rm E-mail: Johannes.Haller@cern.ch}}

\abstract{
The HERA collaborations, H1 and ZEUS, have searched for physics beyond the Standard Model in $e^{\pm}p$ collisions at centre-of-mass energies of up to 319\,GeV.
In this article the experimental results coming from the HERA\,I phase are summarised and the constraints on theories of new physics including Contact Interactions, Large Extra Dimensions, Leptoquarks, FCNC and various scenarios in $R$-parity violating supersymmetry are discussed.
In addition the first results on searches for new physics coming from the upgraded HERA\,II collider are presented.
}

\normalsize\baselineskip=15pt

\section{Introduction}
The Standard Model (SM) of particle physics describes the strong, electromagnetic and weak interactions of elementary particles.
It is remarkably confirmed during the last decades by experimental results in both the low- and high-energy regime, in some cases with an amazing accuracy.
Despite this success the SM remains incomplete and unsatisfactory, since many fundamental facts such as the particle mass spectrum, the quark-lepton symmetry and the $SU(3)_{C}\times SU(2)_L\times U(1)_Y$ structure of the gauge groups are not explained. 
In addition, the SM only offers a partial unification of the electroweak and strong forces, whereas gravity is not included.
The famous stability problem of the Higgs mass at the electroweak scale is not solved in the SM.

This dissatisfaction makes the search for physics beyond the SM a central duty for all experiments at high-energy colliders.
For more than a decade, the HERA experiments, H1 and ZEUS, have contributed to this important quest in a significant manner.
At HERA electrons or positrons of $27.6$\,GeV are collided with protons of $920$\,GeV ($820$\,GeV before 1998), resulting in a centre-of-mass energy $\sqrt{s}$ of $319$\,GeV ($301$\,GeV).
In these collisions $eq$ interactions are probed at highest energy. 
New heavy particles could be directly produced with masses up to $\sqrt{s}$.
New physics at scales much larger than $\sqrt{s}$ may also be observable through indirect signatures.

\begin{figure}[t]
\begin{center}
\includegraphics*[width=5.9cm,height=7.6cm]{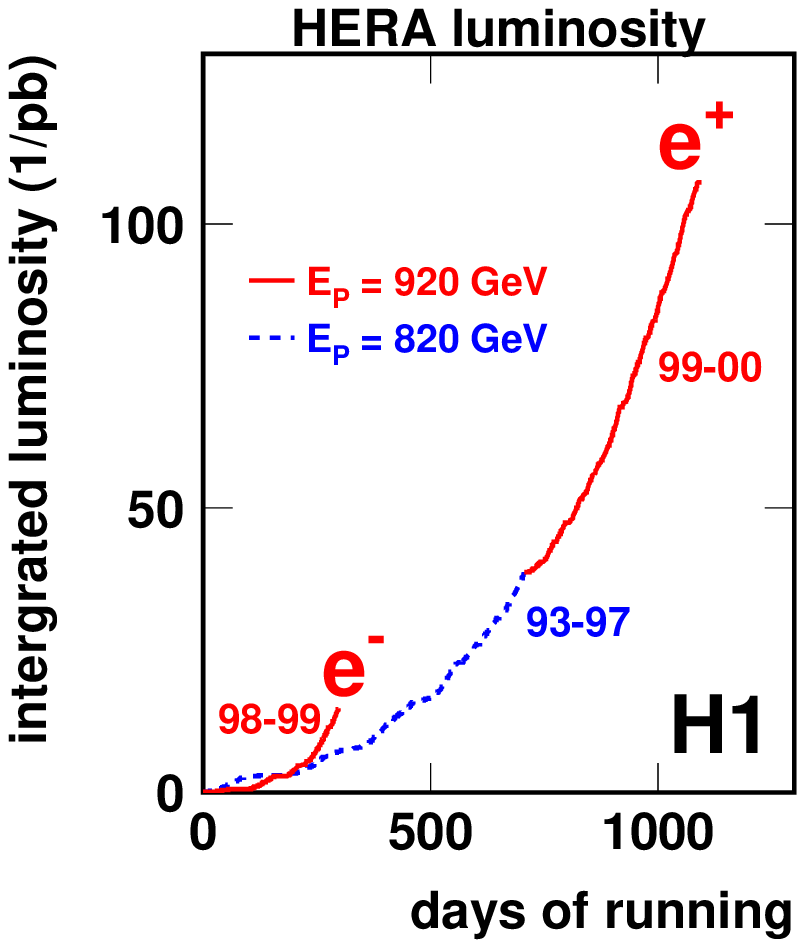}
\includegraphics*[width=8.1cm]{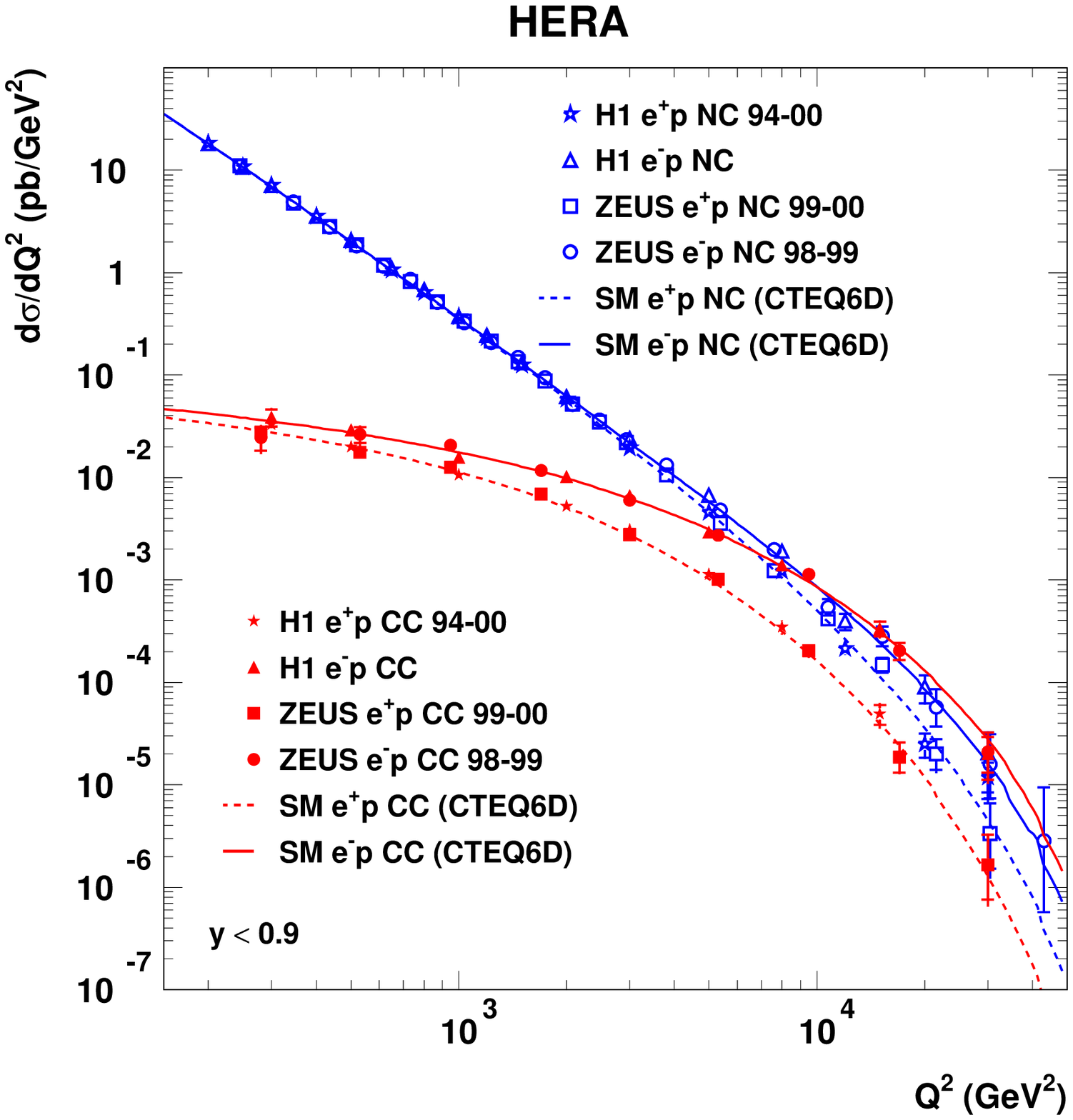}
\caption{ %
(left) Integrated luminosities of the HERA\,I data taking period; (right) Final measurements of the differential cross-sections for NC and CC DIS from HERA\,I.}
 \label{fig:lumi}
\end{center}
\end{figure}

The integrated luminosity collected by H1 during the HERA\,I phase (1992-2000) is illustrated in figure~\ref{fig:lumi}~(left).
It reached roughly $120\,\mathrm{pb}^{-1}$ in $e^+p$ collisions and $20\,\mathrm{pb}^{-1}$ in $e^-p$ collisions.
ZEUS collected a similar amount of data.
The collaborations have searched for deviations from the SM in these data and they have set constraints on various models 
describing new physics beyond the SM. 
In the coming years the HERA collaborations will continue to search for new phenomena with improved experimental devices.
A major upgrade of the HERA collider and its detectors was undertaken in 2000/2001. 
In order to achieve higher luminosities new focussing magnets were inserted close to the two $ep$ interaction points.
Additional longitudinal polarisation of the electron\footnote{In the following the term {\it electron} will be used to refer to both electron and positron unless explicitly stated otherwise.} beams is now provided for the interaction regions by new spin rotators.
The data taking started in 2002/2003. 
The first preliminary results from this HERA\,II period are now available.

\section{Deep Inelastic Scattering at High $\mathbf {Q^2}$}

New phenomena from physics beyond the SM are most likely to appear at high invariant masses, i.e. high energy.
In this kinematic region the dominant SM process at HERA is deep-inelastic $ep$ scattering (DIS) which can be described by the $t$-channel exchange of a gauge boson between the incoming electron and a quark from the proton. 
At low values of the squared four-momentum transfer $Q^2$ only the photon exchange contributes since the exchange of massive gauge bosons ($Z^0,W^{\pm}$) is suppressed by propagator terms.
However, at high $Q^2$ the $Z^0$ and $W^{\pm}$ contributions become important. 
This allows the investigation of electroweak effects in $eq$ interactions.

In figure~\ref{fig:lumi}~(right) the final measurements~\cite{h1nccc,znccc} of the differential neutral current (NC) and charged current (CC) cross sections $d\sigma/dQ^2$ coming from HERA\,I are shown together with the SM expectation.
At low $Q^2$ the suppression of the CC cross section ($W^{\pm}$) with respect to the NC cross section ($\gamma,Z^0$) is clearly visible.
At high $Q^2$ both cross sections are of similar size.
With the statistics collected at HERA\,I, the measurements clearly reveal the dependence on the lepton beam charge that is predicted by the SM, i.e. an increased NC cross section for $e^-p$ with respect to $e^+p$ scattering by virtue of positive ($e^-p$) or negative ($e^+p$) interference between $\gamma$ and $Z^0$ exchange.

\begin{figure}[t]
\begin{center}
\includegraphics*[width=10cm]{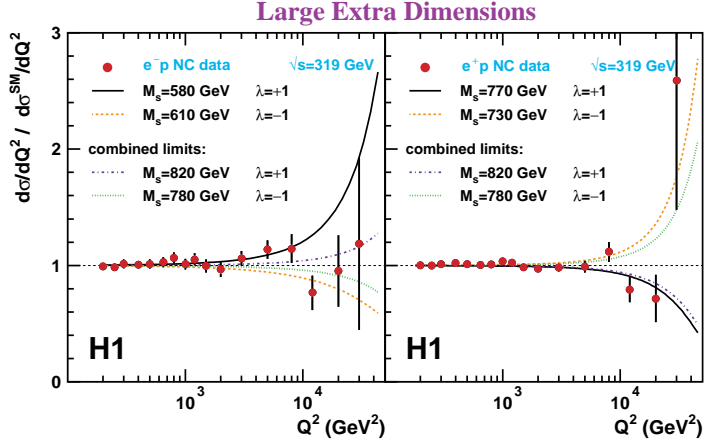}
\caption{%
Ratio of data and SM expectation for the NC cross section as a function of $Q^2$ as measured by H1. The lines illustrate fits of various models with Large Extra Dimensions, which correspond to 95\,\% CL exclusion limits on the effective Planck Mass $M_s$.}
\label{fig:led}
\end{center}
\end{figure}

New physics at scales $\Lambda\gg\sqrt{s}$ could produce deviations of the observed NC DIS cross section from the SM prediction at high $Q^2$.
E.g. the virtual exchange of a new heavy boson could interfere with the $\gamma$ and $Z^0$ fields of the SM.
Such indirect signatures can be interpreted in the formalism of four-fermion point-like contact interactions (CI), where the boson propagator is contracted.
The most general chiral invariant Lagrangian for NC vector-like contact interactions can
 be written in the form~\cite{cilang} 
$
{\cal L}_V=\sum_{a,b=L,R}\eta_{ab}^q(\overline{e}_a\gamma^{\mu
}e_a)(\overline{q}_b\gamma_{\mu}q_b), 
$
where $\eta_{ab}^q=\epsilon\frac{g^2}{(\Lambda^q_{ab})^2}$ are model
 dependent coefficients of the new process, $g$ is the coupling constant, $\Lambda_{ab}^q$ is the effective mass scale and $\epsilon=\pm1$ is a parameter determining the sign of the interference with the SM.   
From fits to the NC DIS measurements at high $Q^2$ performed by both collaborations~\cite{h1ci,zci} limits ($95\,\%$ CL) on the scale $\Lambda$ have been derived.
These limits range up to $7$\,TeV for the various CI models and are comparable to those obtained at LEP and the Tevatron studying reactions complementary to HERA.

Constraints on models with Large Extra Dimensions (LED)~\cite{arkani} can be obtained in a similar way. 
In such models SM particles can propagate in the ordinary 4-dimensional space, whereas gravitons live in a world with $n\ge2$ extra compactified dimensions.
This model solves the hierarchy problem of the SM by introducing an effective Planck scale $M_s$. 
This scale is related to the ordinary Planck scale $M_p$ by 
$M_p^2=R^nM_s^{2+n}$,
where $R$ is the size of the extra dimensions. 
Gravitons are expected to appear in the $(3+1)$-dimensional world as towers of Kaluza-Klein modes.
Gravitation is then a strong force at short distances.
Since gravitation has been tested directly only to the millimetre scale, it is attractive to test LED models at HERA or other collider facilities.
In the LED Lagrangian an additional term $\propto\epsilon/M_s^4$ arises, which accounts for the graviton exchange.
This term leads to deviations from the SM expectation of the measured NC DIS cross section. 
In figure~\ref{fig:led} the NC cross section ratio for data and SM expectation as measured by H1 is shown for the $e^-p$ and $e^+p$ data together with fits in the LED model. 
These fits correspond to $95$\,\% CL exclusion limits on the effective scale $M_s$.
The current HERA limits~\cite{h1ci,zci} are of the order of $0.8$\,TeV.
At high $Q^2$ the measurements are limited by statistics.
Future data from HERA\,II will lead to more precise cross section measurements and to more stringent exclusion bounds.

Similarly, a finite radius of electrons and quarks would lead to deviations of the NC cross section.
Fits to the data~\cite{h1ci,zci} yield upper limits on these radii. 
Assuming a point-like electron the radius of the light $u$ and $d$ quarks is constrained to around $R_q < 0.8 \cdot 10^{-18}$\,m at $95\,\%$ CL by the HERA results.
These limits are similar to limits set by CDF~\cite{cdfci} in $p\bar{p}$ collisions.
L3~\cite{l3ci} has presented a stronger limit ($R_q<0.4\cdot 10^{-18}$\,m) assuming the same radius for all produced quark flavours.

\section{Observation of Outstanding Events with Leptons at HERA\,I}

A general search for new physics performed by H1~\cite{h1gen} investigated many final states with possible sensitivity to new phenomena.
It shows that the data taken during HERA\,I are generally in very good agreement with the SM prediction. 
However, event signatures have been observed which deviate from the SM.
These deviations had already been found in former analyses~\cite{h1isolep98,h1isolep,h1mulel,zmullep}.

\subsection {Events with Isolated Charged Leptons and Missing Transverse Momentum}

\begin{figure}[t] 

\begin{center}

\includegraphics*[width=5.9cm]{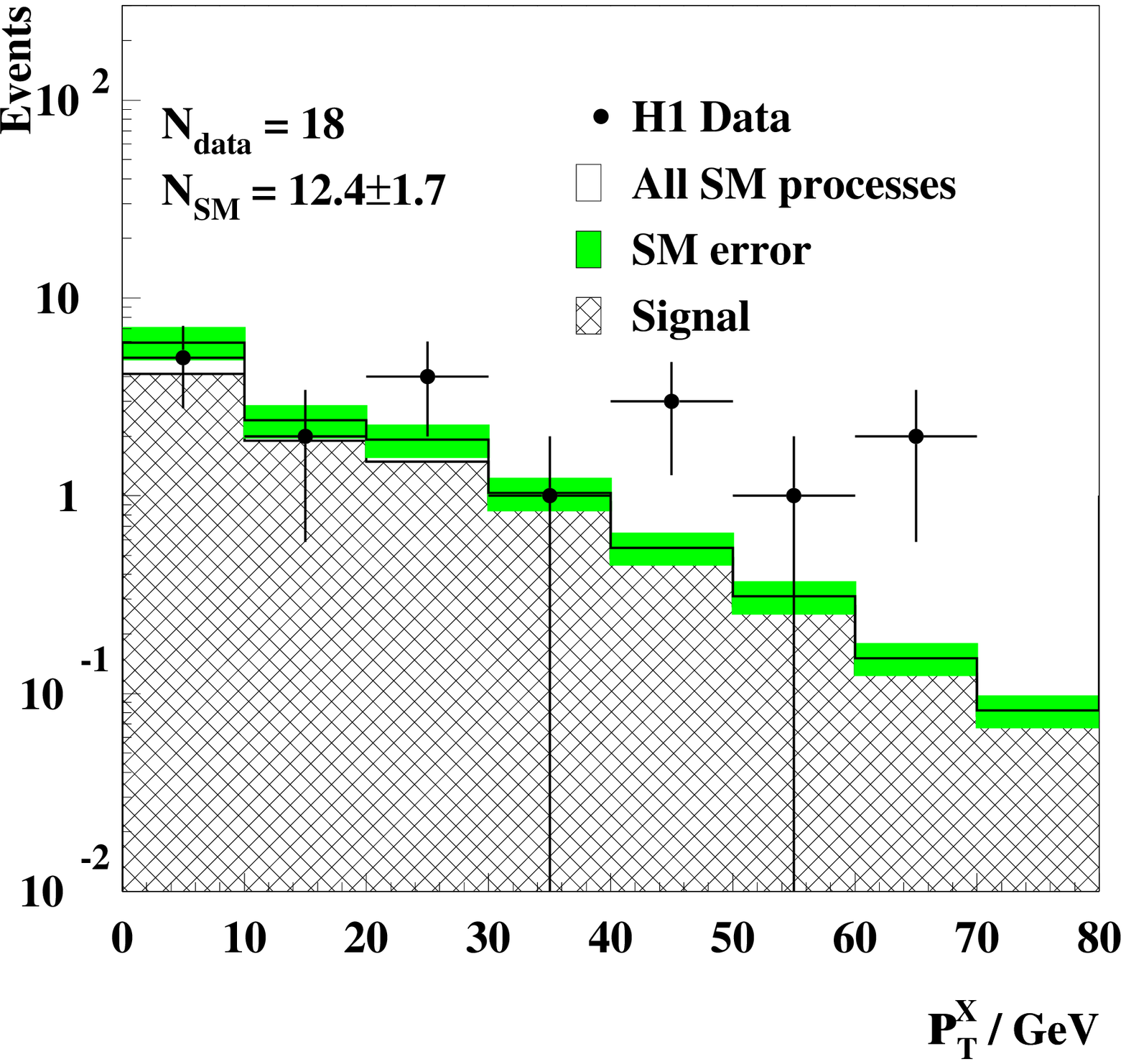}
\includegraphics*[width=8.9cm]{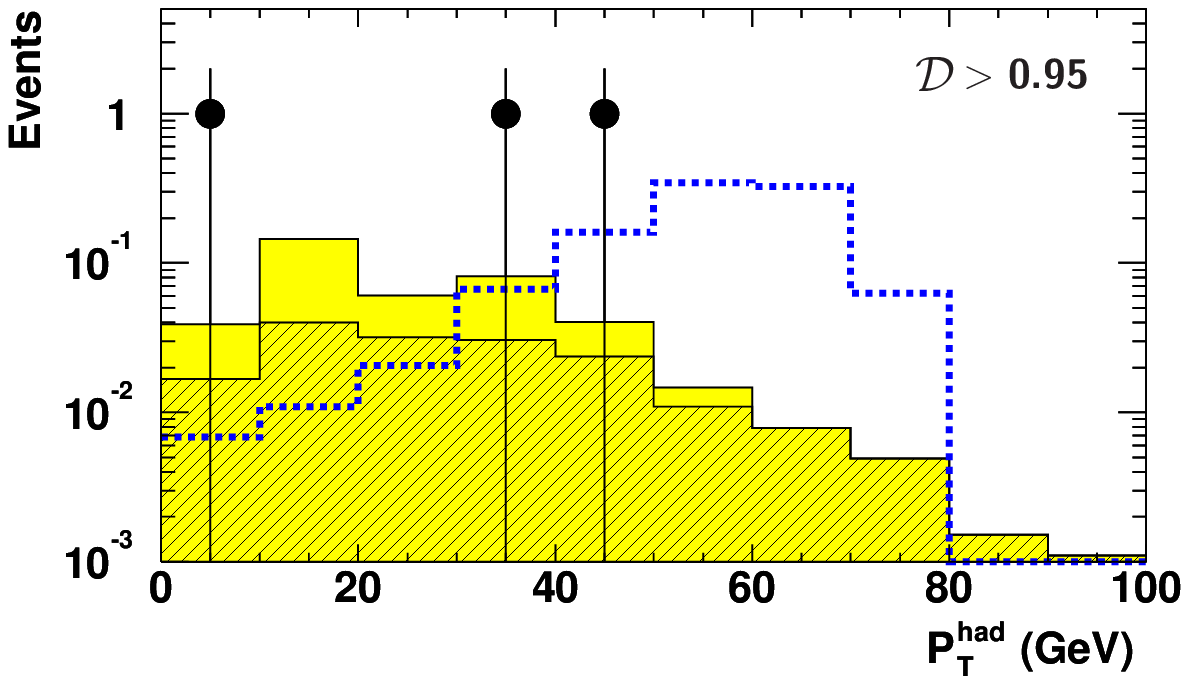}

\caption{%
(left) Distribution of events with missing transverse momentum and isolated electrons or muons as a function of $p_T^{\rm had}(=p_T^{\rm X})$, the transverse momentum of the hadronic system, as measured by H1 in the HERA\,I $e^+p$ dataset; (right) Distribution of events with isolated tau leptons selected by the ZEUS experiments in the full $e^{\pm} p$ dataset. The SM expectation (shaded) and the distribution for FCNC single top production (dashed) are also shown.} 
   \label{fig:isolep}
  \end{center}
\begin{picture}(0,0)
\put(375,180){\bf \Large ZEUS}
\end{picture}
\end{figure}

In 1998  H1 reported an excess of striking events with isolated leptons (electrons or muons), large missing transverse momentum and large transverse momentum of the hadronic final state $p_T^{\rm had}$~\cite{h1isolep98}.
This event topology has also been observed in the full HERA\,I dataset~\cite{h1isolep}.
Within the SM such events are expected mainly due to $W$ boson production with subsequent leptonic decay.
Without a cut on $p_T^{\rm had}$, in $e^-p$ interactions one event is observed in the electron channel and none in the muon channel, consistent with the expectation of the SM.
In the $e^+p$ data a total of 18 events are seen in the electron and muon channels compared to an expectation of $12.4\pm 1.7$ dominated by $W$ production which is known in NLO QCD~\cite{wprod}.
Whilst the overall observed number of events is broadly in agreement with the predicted number, there is an excess of events in the $e^+p$ data with  $p_T^{\rm had}>25\,\GeV$ with 10 events found compared to $2.9\pm 0.5$ expected.
In figure~\ref{fig:isolep}~(left) the distribution in $p_T^{\rm had}$ of these events is shown for the $e^+p$ data set alone.
A similar analysis, performed by the ZEUS Collaboration found no deviation from the SM in both $e^+p$ and $e^-p$ collisions~\cite{ztop}.
The total event numbers of the two experiments for the full $e^{\pm}p$ dataset are summarised in table~\ref{tab:iltotnum} for two different $p_T^{\rm had}$ selection cuts. 

\begin{table}[b]
 \begin{center}
  \caption{Total event numbers in the full HERA\,I $e^{\pm}p$ dataset and SM expectations for events with isolated leptons and missing transverse momentum for two different cuts on $p_{T}^{\rm had}$.} 
  \begin{tabular}{|c|c|c|c|}
  \hline
  H1 & electron & muon & tau (prelim.) \\ \hline
  $p_{T}^{\rm had}>25 \GeV$ & 5 (1.8$\pm$0.3) & 6 (1.7$\pm$0.3) & 0 (0.53$\pm$0.10) \\ 
  $p_{T}^{\rm had}>40 \GeV$ & 3 (0.7$\pm$0.1) & 3 (0.6$\pm$0.1) & 0 (0.22$\pm$0.05) \\ \hline\hline
  ZEUS & electron & muon & tau \\ \hline
  $p_{T}^{\rm had}>25 \GeV$ & 2 (2.9$\pm$0.6) & 5 (2.8$\pm$0.2) & 2 (0.20$\pm$0.05) \\ 
  $p_{T}^{\rm had}>40 \GeV$ & 0 (0.9$\pm$0.1) & 0 (0.9$\pm$0.1) & 1 (0.07$\pm$0.02) \\ \hline
 \end{tabular}
  \label{tab:iltotnum}
 \end{center}
\end{table}

The two collaborations also investigated the corresponding event signature with isolated tau leptons.
ZEUS performed a search for isolated tracks associated with pencil-like hadronic jets~\cite{ztau}, selected by using observables based on the internal jet structure to discriminate between tau decays and quark- or gluon-induced jets. 
The distribution of selected events in $p_T^{\rm had}$ is shown in figure~\ref{fig:isolep}~(right). 
Two events are found with $p_T^{\rm had}>25\,\GeV$, where only $0.20\pm0.05$ events are expected.
The total event numbers together with the SM expectation are summarised in table~\ref{tab:iltotnum}. 
A preliminary H1 search for isolated tau leptons shows no deviation from the SM~\cite{h1tau}.

More data are needed to fully clarify the situation of these striking events.
The first results of analyses of data from HERA\,II are discussed in section 6.

The production of a single top quark decaying via $t\rightarrow bW$ would naturally lead to the observed event topology.
At HERA, SM single top production at the tree level can only proceed via a CC reaction $ep\rightarrow \nu t \bar{b}X$.
At HERA energies this process has a cross section of less than 1\,fb, too small to be detected with the present accumulated luminosities. 
Flavour changing neutral currents (FCNC) processes are only present via higher order radiative corrections and are highly suppressed by the GIM mechanism.
However, in many extensions of the SM, the presence of anomalous FCNC vertices $tu\gamma$ and $tuZ$ leads to the NC reaction $ep\rightarrow etX$.
This FCNC transition can be parameterised by an effective Lagrangian~\cite{toptheo} in which the magnetic coupling $\kappa_{tu\gamma}$ and the vector coupling $v_{tuZ}$ are usually considered.

\begin{figure}[t] 
\begin{center}
\includegraphics*[width=5.9cm]{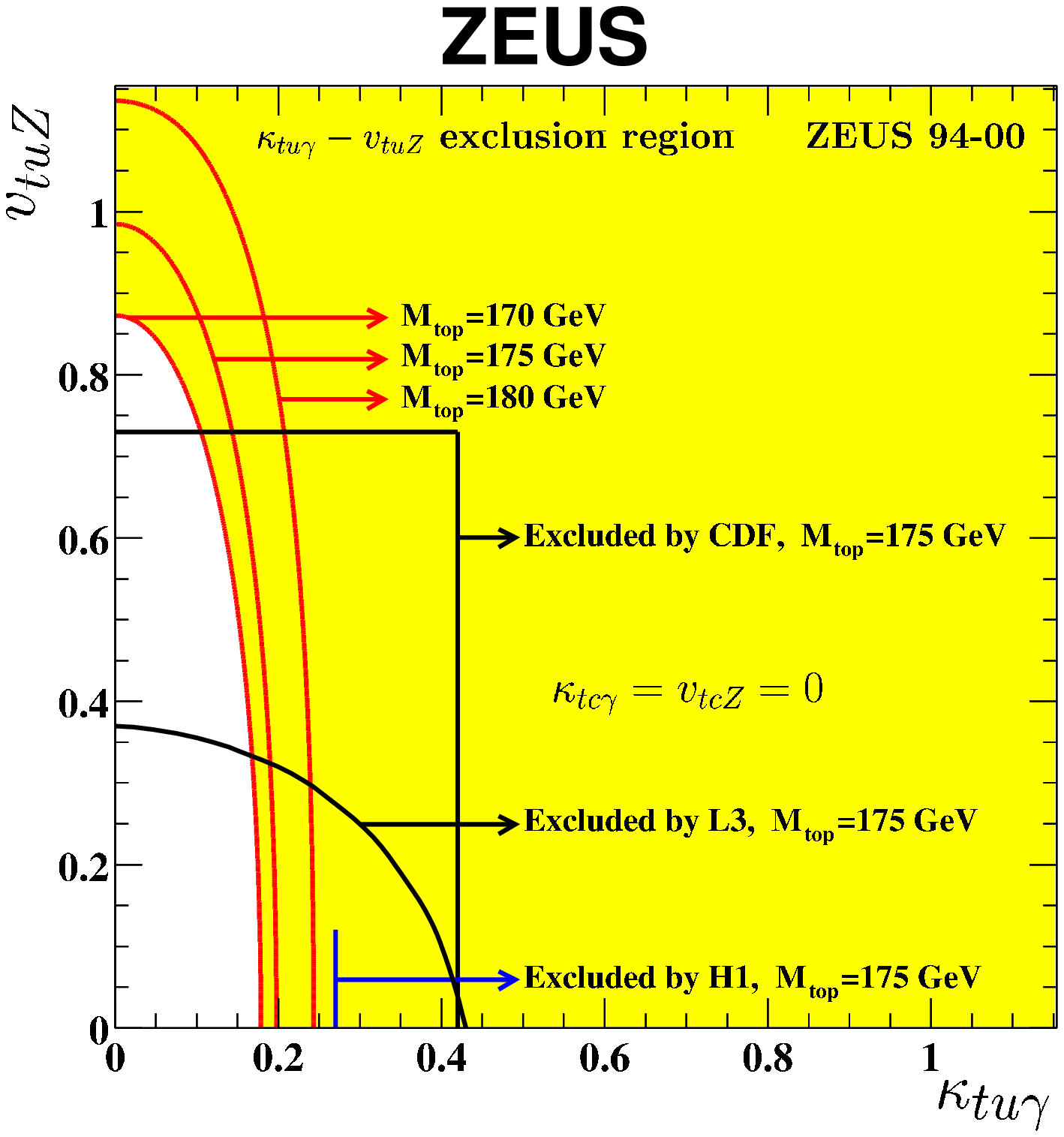}
\caption{
Constraints ($95\,\%$ CL) on anomalous single top quark production: Exclusion area in the plane spanned by the FCNC couplings $\kappa_{tu\gamma}$ and $v_{tuZ}$.}
\label{fig:top}
\end{center}
\end{figure}

H1 performed a dedicated analysis~\cite{h1top} with selection cuts optimised for single top-quark events. 
In the leptonic decay channel of the $W$ from the top decay, five events were observed while $1.3\pm0.2$ were expected from the SM.
ZEUS observed no event that could be a candidate for anomalous single top-quark production~\cite{ztop}. 
Also the top decay to hadrons has been investigated by both collaborations. 
The results are in agreement with the SM. 
After the combination of leptonic and hadronic channels, limits on the anomalous couplings $\kappa_{tu\gamma}$ and $v_{tuZ}$ can be derived.
The excluded area in the ($\kappa_{tu\gamma},v_{tuZ}$)--plane is shown in figure~\ref{fig:top}. 
The HERA results are mainly sensitive to $\kappa_{tu\gamma}$ and exclude a significant area not previously excluded.
H1 sets less stringent limits than ZEUS because of the excess in the leptonic channel.

\subsection{Events with Multiple Leptons at High Transverse Momentum}

\begin{figure}[t] 
\begin{center}
\includegraphics*[width=5.5cm,clip=]{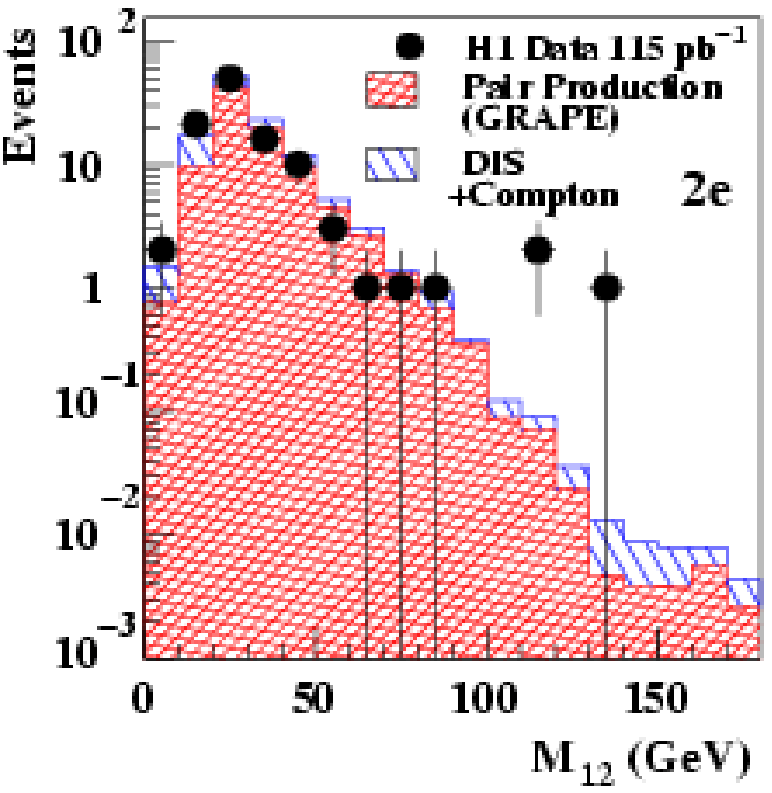}
\includegraphics*[width=5.5cm,clip=]{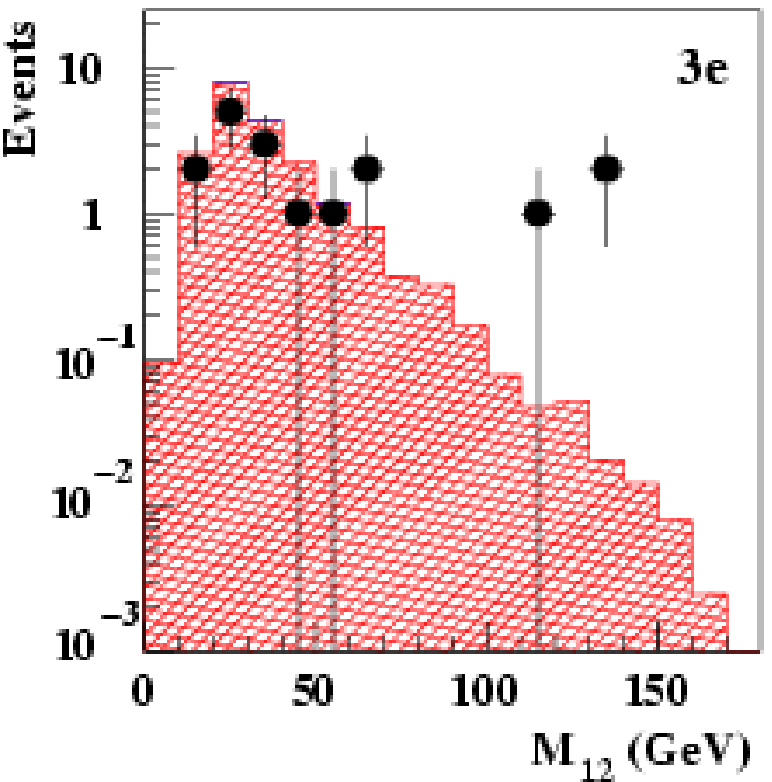}
\caption{Distribution of events with two (left) and three (right) electrons of high $p_T$ as a function of the invariant mass $M_{12}$ of the two highest $p_T$ electrons compared with the SM expectation.}
\label{fig:mulel}    
\end{center}
\end{figure}

The H1 Collaboration also reported a measurement of multi-electron production at high $p_T$~\cite{h1mulel}. 
Figure~\ref{fig:mulel} shows the distribution of selected events with two and three electrons in the final state as a function of the invariant mass $M_{12}$ of the two highest $p_T$ electrons as well as the SM expectation.
Six events were observed with a di-electron mass above 100\,GeV, a domain where the SM prediction is low ($0.53\pm0.06$).
In this mass region the ZEUS experiment found two events with a SM expectation of $1.2\pm0.1$ in a preliminary analysis~\cite{zmullep}.
A measurement of multi-muon production in H1~\cite{h1mulmu} showed good agreement between the data and the SM expectation over the whole mass range.
The new data from HERA\,II are needed for clarification (cf. section 6).

One possible explanation for the observation of events with multiple electrons at high $p_T$ is the production of doubly-charged Higgs bosons.
These appear in various extensions of the SM, in which the usual Higgs sector is extended by one or more triplet(s) with non-zero hypercharge.
Examples are provided by some Left-Right symmetric (LRS) models~\cite{lrs}.
At the tree level, doubly-charged Higgs bosons couple only to charged leptons and to other Higgs and gauge bosons.
HERA allows for single production of doubly-charged scalars via a $h_{ee}$ coupling by the fusion of the incoming beam electron with an electron emerging from the splitting of a photon radiated off the proton.
When only diagonal couplings are considered, the reaction $e^{\pm}p\rightarrow e^{\mp}H^{\pm\pm}X$ followed by the decays $H^{\pm\pm}\rightarrow e^{\pm}e^{\pm} (\mu^{\pm}\mu^{\pm},\tau^{\pm}\tau^{\pm})$ could lead to the event topology in question.
A dedicated analysis has been performed by the H1 Collaboration~\cite{h1dch}.
When kinematic cuts and lepton charges are taken into account, only one of the multi-electron events was found to be compatible with the hypothesis of the decay of a heavy Higgs boson.
Assuming that the doubly-charged Higgs only decays to electrons, a lower limit of about 130\,GeV on the $H^{\pm\pm}$ mass is set for $h_{ee}=0.3$.

\section{Searches for Leptoquarks}

\begin{figure}[t]
\begin{center}
\includegraphics*[width=7.0cm,clip=]{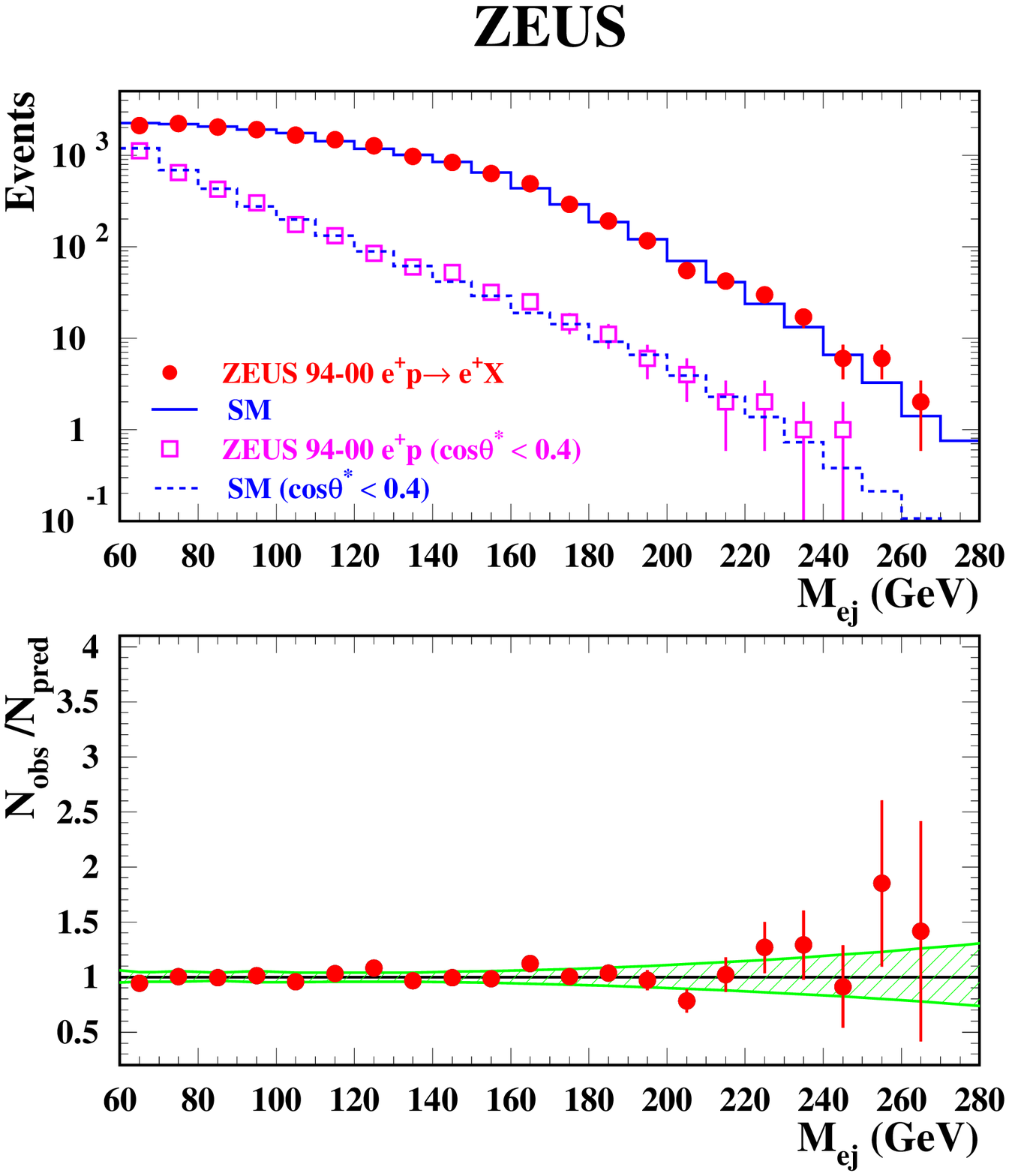}
\includegraphics*[width=7.0cm,clip=]{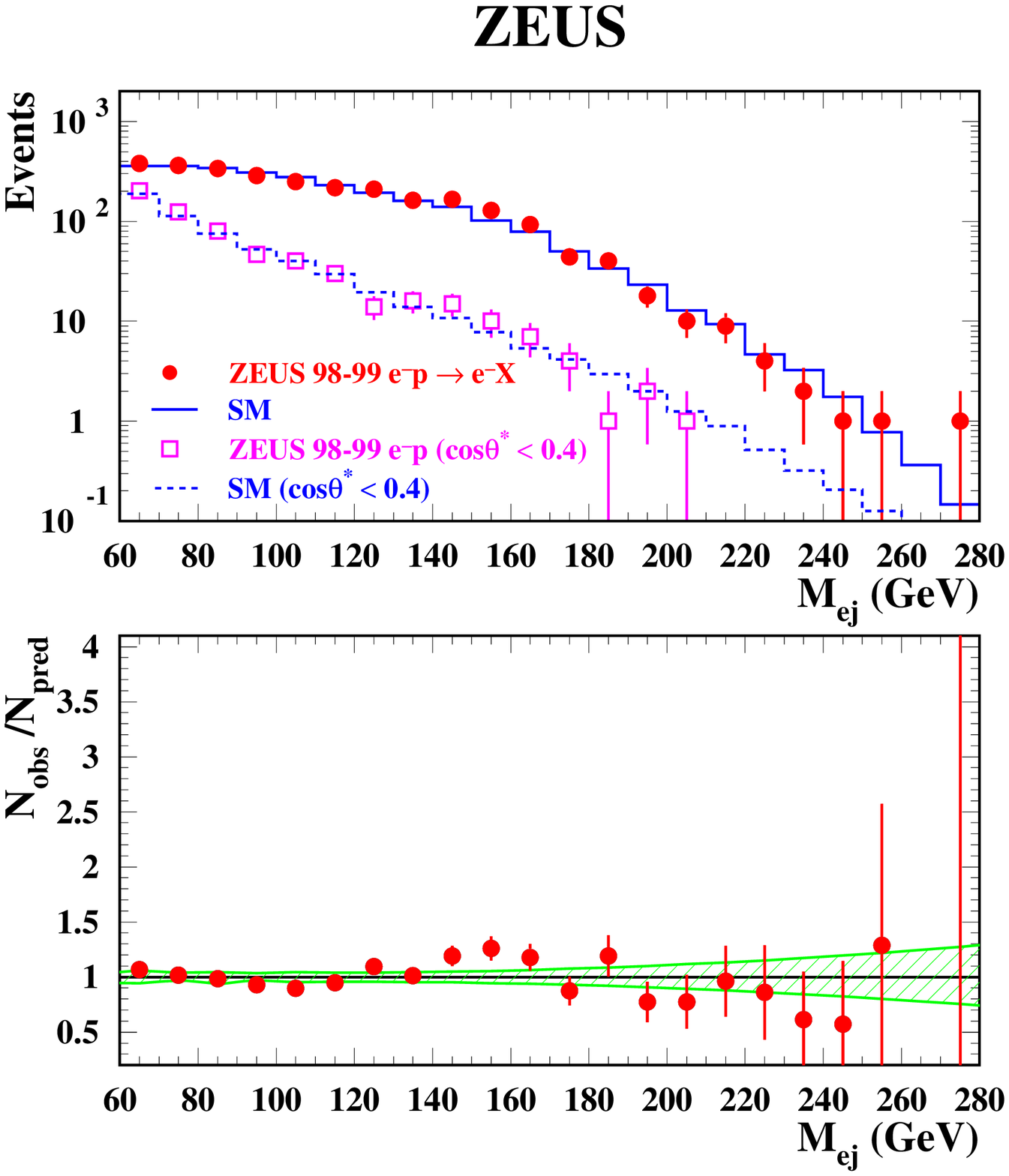}
\caption{ Invariant $e+$jet mass spectra in $e^+p$ and $e^-p$ collisions as measured by ZEUS.}
  \label{fig:lqspec}
\end{center}
\end{figure}

\begin{figure}[t]
\begin{center}
\includegraphics*[width=7.cm,clip=]{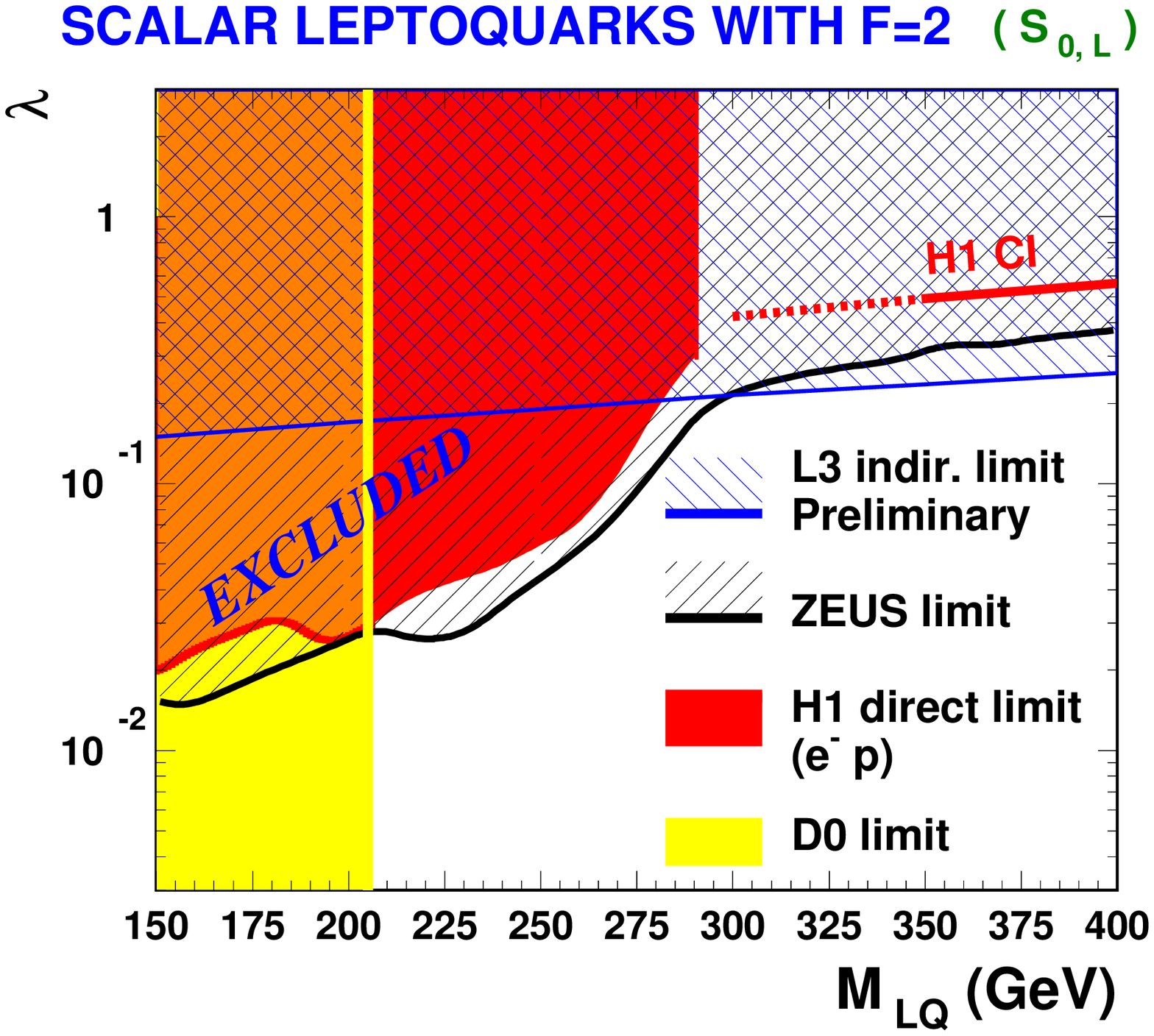}
\includegraphics*[width=7.cm,clip=]{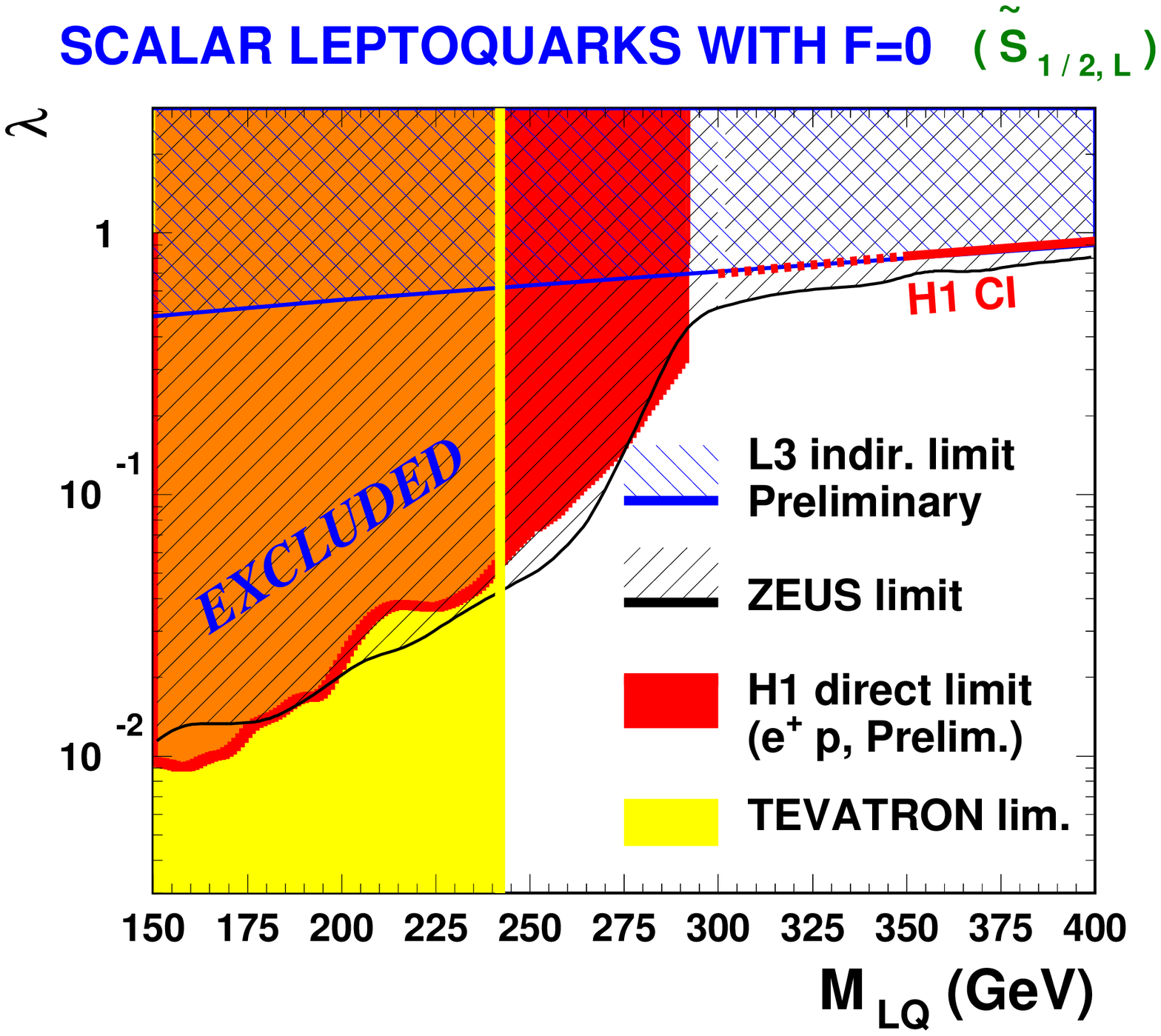}
\caption{ Constraints ($95\,\%$ CL) on the leptoquark coupling $\lambda$ as a function of $M_{\rm LQ}$ in the framework of the BRW model for a leptoquark with fermion number $F=2$ (left) and $F=0$ (right).}
\label{fig:lqlim}
\end{center}
\end{figure}


Leptoquarks (LQ) are colour triplet bosons of spin $0$ or $1$, carrying lepton ($L$) and baryon ($B$) number and fractional electric charge. They couple to lepton-quark pairs and appear in many extensions of the SM which unify leptons and quarks in the framework of Grand Unified Theories (GUT), e.g. superstring-inspired $E_6$ models, Technicolour-like theories and Compositeness models. 

LQs with masses $M_{\rm LQ}<\sqrt{s}$ could be directly produced at HERA by fusion of the initial state electron and a quark from the incoming proton. 
The $s$-channel production cross section depends on the unknown Yukawa coupling $\lambda$ at the electron-quark-leptoquark vertex and on the LQ mass.
It is approximately given by $\sigma(ep) \propto {\lambda}^2 \cdot q \left( x=M_{\rm LQ}^2/s\right)$ where $q(x)$ is the density of the struck quark in the incoming proton.
The $u$-channel exchange of a LQ and its interference with the corresponding SM process gives access to masses above the centre-of-mass energy of HERA.

The most general effective LQ model was proposed by Buchm\"uller, R\"uckl and Wyler (BRW)~\cite{brw} under the assumptions that i) LQs have renormalisable interactions which are invariant under the SM gauge groups, ii) LQs couple only to SM fermions and gauge bosons and iii) LQs conserve $L$ and $B$ separately.
In this model 14 possible scalar or vector LQs with fermionic number $F=-(3B+L)=0$ or $2$ appear.
Their branching ratios $\beta_{eq}$ and $\beta_{\nu q}$ to $eq$ and $\nu q$ have fixed values of 1, $\frac{1}{2}$ or 0.
At HERA LQs with $F=0$, which couple simultaneously to one fermion and one antifermion, are better probed in $e^+p$ collisions due to the large valence quark densities of the proton. 
In contrast, the seven $F=2$ LQs, which couple to two fermions or two antifermions, are better probed in $e^-p$ collisions.

The final states of LQ decays and SM DIS reactions are the same ($e+$\,jet and $\nu+$\,jet).
The polar angular distribution of the events though is different.
Scalar LQs produced in the $s$-channel decay isotropically in their rest frame leading to a flat $d\sigma/dy$ distribution, where $y=\frac{1}{2}(1+\cos\theta^*)$ is the Bjorken scattering variable ({\it inelasticity}) and $\theta^*$ is the polar angle of the lepton in the LQ centre-of-mass frame. 
In contrast, for vector LQ production the events  would be distributed according to  $d\sigma/dy\propto(1-y)^2$. Both $d\sigma/dy$ spectra are markedly different from the $d\sigma/dy\propto y^{-2}$ distribution expected for the dominant $t$-channel photon exchange in NC DIS events.
These differences are exploited by the HERA collaborations to enhance the LQ signal over the DIS background.

The $s$-channel production of LQs would lead to a resonance peak in the invariant mass distribution of the decay products.
The invariant mass spectra for the $eq$ final state measured by ZEUS~\cite{zlq} are shown in figure~\ref{fig:lqspec} for the full $e^+p$ and $e^-p$ dataset of HERA\,I.
No evidence of a signal has been found by either collaboration~\cite{zlq,h1lq}. 
Thus the excess observed in NC DIS $e^+p$ data of 1994-1997 at invariant masses of about $200$\,GeV is not confirmed by the full dataset.

Both collaborations have derived limits on the Yukawa coupling $\lambda$ as a function of the LQ mass for all LQ types in the BRW model. 
The results for a typical scalar LQ with $F=2$ ($S_{0,L}$) for which $\beta_{eq}=0.5$ and for a scalar LQ with $F=0$ ($\tilde{S}_{1/2,L}$) for which $\beta_{eq}=1$ are shown in figure~\ref{fig:lqlim}. 
The HERA limits are particularly stringent for masses up to the centre-of-mass energy where the LQ can be produced directly. 
At higher masses the $u$-channel LQ exchange gives sensitivity.
For comparison, the indirect limits obtained from the hadronic cross section in $e^+e^-$ collisions at LEP and the results from searches for direct LQ pair production at Tevatron are also shown.

In more general LQ models still without lepton flavour violation (LFV) the branching ratios $\beta_{eq}$ and $\beta_{\nu q}$ are treated as free parameters assuming $\beta_{eq}+ \beta_{\nu q}=1$.
In this case HERA limits obtained by combining the NC and CC channels are almost independent of the branching ratios.
The comparison with Tevatron limits shows that the HERA results are particularly stringent for small $\beta_{eq}$ where the Tevatron experiments suffer from large QCD background.
HERA results on LQ models with LFV can be found in~\cite{h1lfv} and \cite{zlfv}.

\section{Supersymmetry with $\mathbf{R_p}$-Violation}

Supersymmetry (SUSY) is likely to be an essential property of a theory beyond the SM.
Among the most compelling arguments for SUSY are the stability of a softly broken SUSY, which naturally avoids arbitrary fine tuning of parameters and the explanation of the hierarchy between the GUT scale (or the Planck scale) and the electroweak mass scale.

In the SM, the conservation of $B$ and $L$ is an automatic consequence of the gauge invariance and renormalisability.
In contrast, $B$ and $L$ are no longer protected in SUSY.
Interactions violating $B$ or $L$ can be avoided {\it ad hoc} by introducing a discrete symmetry implying the conservation of $R$-parity, defined as $R_p=(-1)^{3B+L+2S}$, where $S$ denotes the particle spin.
Hence $R_p$ is equal to $+1$ for particles and equal to $-1$ for sparticles. 
Whether or not $R_p$ is conserved has dramatic consequences for SUSY searches. 
If $R_p$ is conserved, sparticles can only be produced in pairs and the lightest supersymmetric particle (LSP) is stable. 
In such frameworks the LSP is a natural candidate for Cold Dark Matter in cosmology and leads to the golden signature of missing energy in collider experiments.

Searches for minimal SUSY at HERA with conservation of $R_p$ have been performed by both the H1 and ZEUS experiments~\cite{h1rpc,zrpc}, looking for the production of a selectron-squark pair.
HERA's sensitivity in such SUSY frameworks is however limited, when taking into account the mass range allowed by other experiments.

However, the most general renormalisable SUSY theory which preserves the gauge invariance of the SM violates $R_p$.
Its superpotential, where only the trilinear terms are considered, is given by
\begin{equation}
\label{eqn:syspotrpviol}
W_{R\!\!\!/}=\frac{1}{2}\lambda_{ijk}L_iL_j\overline{E}_k+\lambda
'_{ijk}L_iQ_j\overline{D}_k + \frac{1}{2}\lambda''_{ijk}\overline
{U}_i\overline{D}_j\overline{D}_k ,
\label{eq:rpv}
\end{equation}
where $i,j$ and $k$ are family indices and $\lambda,\lambda'$ and $\lambda''$ are $R_p$-violating (\Rp) Yukawa couplings.
$L$, $E$, $Q$, $D$ and $U$ are superfields, which contain the left-handed leptons, the right-handed electron, the left-handed quarks, the right-handed down quark and the right-handed up quark, respectively, together with their SUSY partners $\tilde{l}_L$, $\tilde{e}_R$, $\tilde{q}_L^j$, $\tilde{d}_R$ and $\tilde{u}_R$.
The corresponding couplings of two known SM fermions with a squark or a slepton allow for resonant $s$-channel production of sparticles at colliders. 
The presence of \Rp\ couplings also opens new decay modes for sparticles.
In particular the LSP is not stable anymore as it decays to ordinary SM particles leading to multi-lepton and multi-jet (MJ) final states.

\subsection{Searches for Squarks in \Rp\ Supersymmetry}
\begin{figure}[t]
 \begin{center}
\psfrag{a}[][][0.8][0]{$e^-$}
\psfrag{b}[][][0.8][0]{$u$}
\psfrag{c}[][][0.8][0]{$\tilde{d}_R^k$}
\psfrag{d}[][][0.8][0]{\quad $e^-,\nu_e$}
\psfrag{e}[][][0.8][0]{\quad $u,d$}
\psfrag{l1}[][][0.8][0]{$\lambda'_{11k}$}
\psfrag{l2}[][][0.8][0]{$\lambda'_{11k}$}
\psfrag{N}[][][0.8][0]{(a)}
\includegraphics*[scale=0.4]{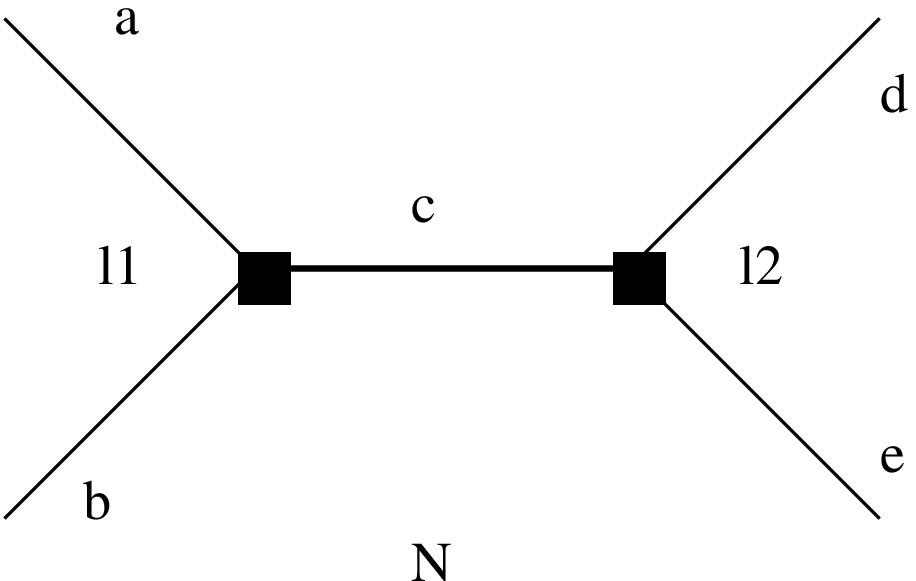}
\psfrag{a}[][][0.8][0]{$e^+$}
\psfrag{b}[][][0.8][0]{$d$}
\psfrag{c}[][][0.8][0]{$\tilde{u}_L^j$}
\psfrag{d}[][][0.8][0]{$u,d$}
\psfrag{e}[][][0.8][0]{$\chi_i^0,\chi^+_i,\tilde{g}$}
\psfrag{f}[][][0.8][0]{$\chi_1^+$}
\psfrag{1}[][][0.8][0]{$eq\bar{q}$}
\psfrag{2}[][][0.8][0]{$(\nu_e q \bar{q})$}
\psfrag{g}[][][0.8][0]{$\chi_1^0$}
\psfrag{j}[][][0.8][0]{$W^+$}
\psfrag{h}[][][0.8][0]{$q',l^+$}
\psfrag{i}[][][0.8][0]{$\bar{q},\nu_l$}
\psfrag{l1}[][][0.7][0]{$\lambda'_{1j1}$}
\psfrag{l2}[][][0.7][0]{$\lambda'_{1j1}$}
\psfrag{N}[][][0.8][0]{(b)}
\includegraphics*[scale=0.4]{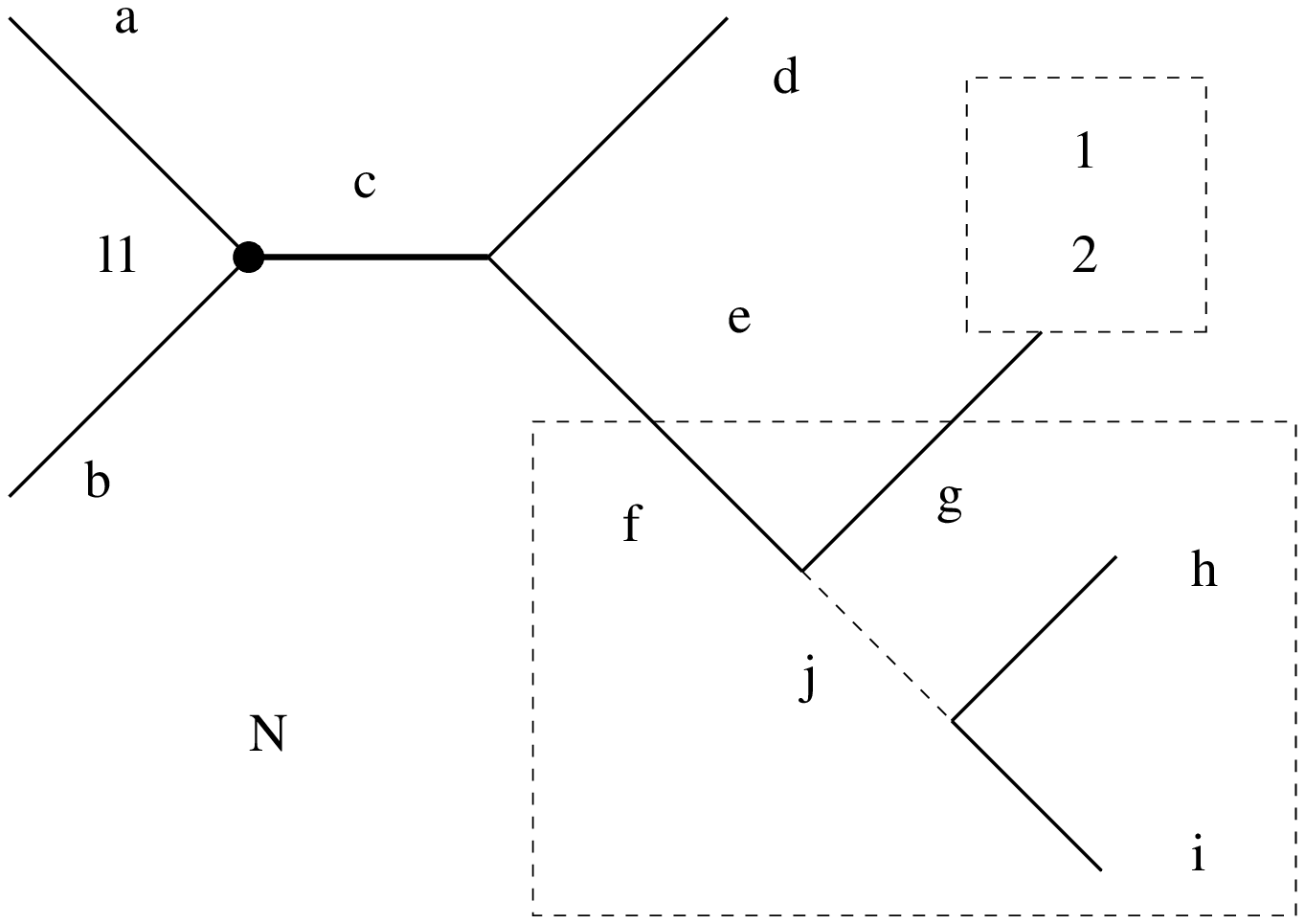}
 \caption{Examples for resonant production and decay of squarks in \Rp\ Supersymmetry with non-vanishing $\lambda'$: The squark can decay (a) violating $R_p$ directly or (b) in a cascade decay to the LSP which subsequently decays violating $R_p$. }
\label{fig:sqdec}
\end{center}
\end{figure}
Of special interest at HERA are couplings between a squark and a lepton-quark pair~\cite{dreiner} as described by the $\lambda'_{ijk}L_iQ_j\overline{D}_k$ term in equation~\ref{eq:rpv}.
A non-zero value of $\lambda'_{1jk}$ would imply the possibility to produce a single squark via $eq$ fusion in the $s$-channel.
This production mechanism offers the best discovery reach of SUSY at HERA. 
The production cross section of squarks is identical to those of leptoquarks and depends on the coupling $\lambda'$ and on the squark mass.
With $e^{\pm}p$ collisions all flavours of squarks can be probed: $e^+p$ collisions give access to up-type squarks via $\lambda'_{1j1}$, whereas $e^-p$ collisions probe the production of down-type squarks via $\lambda'_{11k}$.

The squark can decay via two types of processes. 
First, it can decay violating $R_p$ directly into an electron or neutrino and a quark. 
In the second mode the squark decays first into a quark and a gaugino ($\chi_i^0,\chi_i^{\pm},\tilde{g}$). 
The latter decays violating $R_p$ to a lepton and two quarks or to a SM gauge boson and a lighter gaugino which in turn decays violating $R_p$.
These decay mechanisms lead to a large variety of possible final state topologies, among them very striking signatures with multiple leptons or a lepton with charge opposite to the initial lepton beam's charge. 
Example diagrams for production and decay of squarks are illustrated in figure~\ref{fig:sqdec}.

\begin{table}[t]
 \begin{center}
  \caption{Total numbers of selected events and SM expectations of squark decay channels in \Rp\ SUSY as measured by H1. }
  \begin{tabular}{|c|cc|cc|}
  \hline
  \rule[-2mm]{0mm}{7mm}{\Large H1} & \multicolumn{2}{c|}{$e^+p$ collisions} & \multicolumn{2}{c|}{$e^-p$ collisions} \\  
\multicolumn{1}{|c|}{Channel} & \multicolumn{1}{c}{Data} & {SM expectation} & {Data} & {SM expectation} \\ \hline \hline
  {{$eq$}} & 632 & 628\,$\pm$\,46 & 204 & 192\,$\pm$\,14 \\
  {{$\nu q$}} & ---& --- & 261  & 269\,$\pm$\,21 \\\hline
  {{$eMJ$ (``right'' charge)}} & 72 & 67.5\,$\pm$\,9.5  & 20 & 17.9\,$\pm$\,2.4 \\
  {{$eMJ$ (``wrong'' charge) }} & 0 & 0.20\,$\pm$\,0.14 & 0 & 0.06\,$\pm$\,0.02 \\
  {{$ee MJ$}} & 0 & 0.91\,$\pm$\,0.51   & 0 & 0.13\,$\pm$\,0.03 \\
  {{$e\mu MJ$}} & 0 & 0.91\,$\pm$\,0.38 & 0 & 0.20\,$\pm$\,0.04 \\
  {{$\nu eMJ$}} & 0 & 0.74\,$\pm$\,0.26& 0 & 0.21\,$\pm$\,0.07 \\\hline
  {{$\nu MJ$}} & 30 & 24.3\,$\pm$\,3.6 & 12 & 10.1\,$\pm$\,1.4 \\
  {{$\nu\mu MJ$}} & 0 & 0.61\,$\pm$\,0.12& 0 & 0.16\,$\pm$\,0.03 \\\hline
 \end{tabular}
 \end{center}

  \label{tab:sqtotnum}
\end{table}

\begin{figure}[b]
\begin{center}
\includegraphics*[scale=0.5]{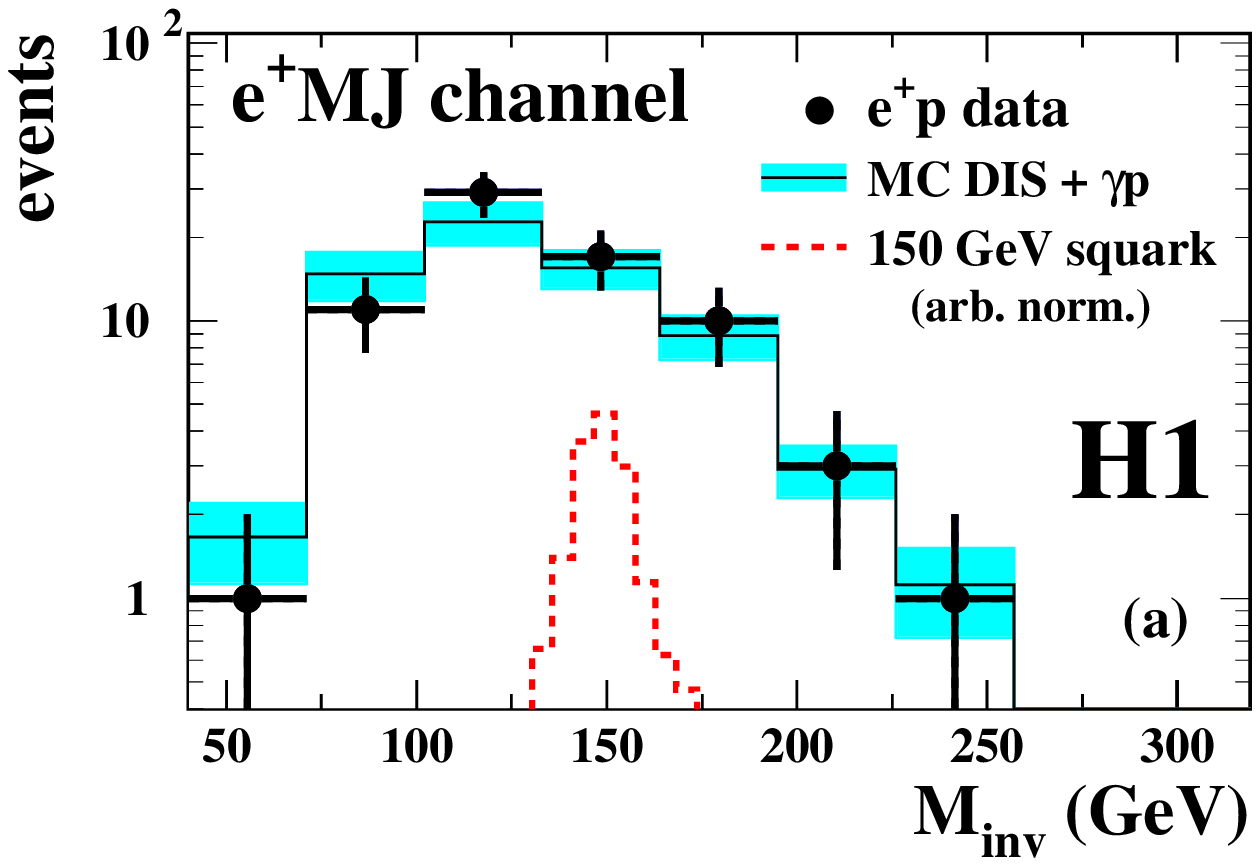}
\includegraphics*[scale=0.5]{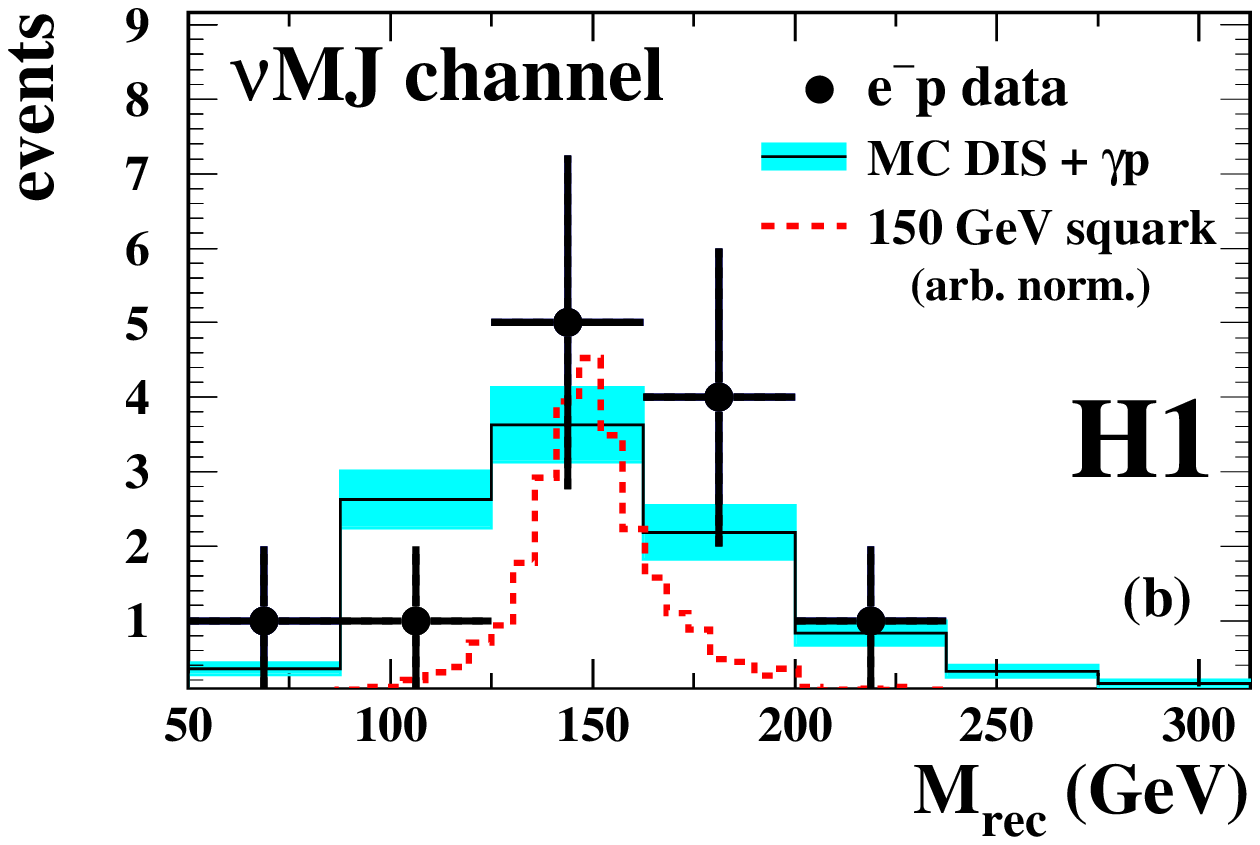}
\caption{ Examples of invariant mass spectra of squark decay channels: (a) the $e^+MJ$ final state 
in $e^+p$ collisions  and (b) the $\nu MJ$ final state in $e^-p$ collisions.}
\label{fig:sqspectra}
\end{center}
\end{figure}

The branching ratios of the various decay channels strongly depend on the SUSY parameters.
In order to be widely independent of these parameters, the search for squark production, as performed by the H1 Collaboration~\cite{h1rpv}, considers all possible final states. 
The total number of selected events and the SM prediction for the various final states are listed in table 2.
Examples of mass spectra are shown in figure~\ref{fig:sqspectra}.
No significant deviation from the SM expectation has been observed in any channel.

\begin{figure}[t] 
\begin{center}
\includegraphics*[width=5.9cm]{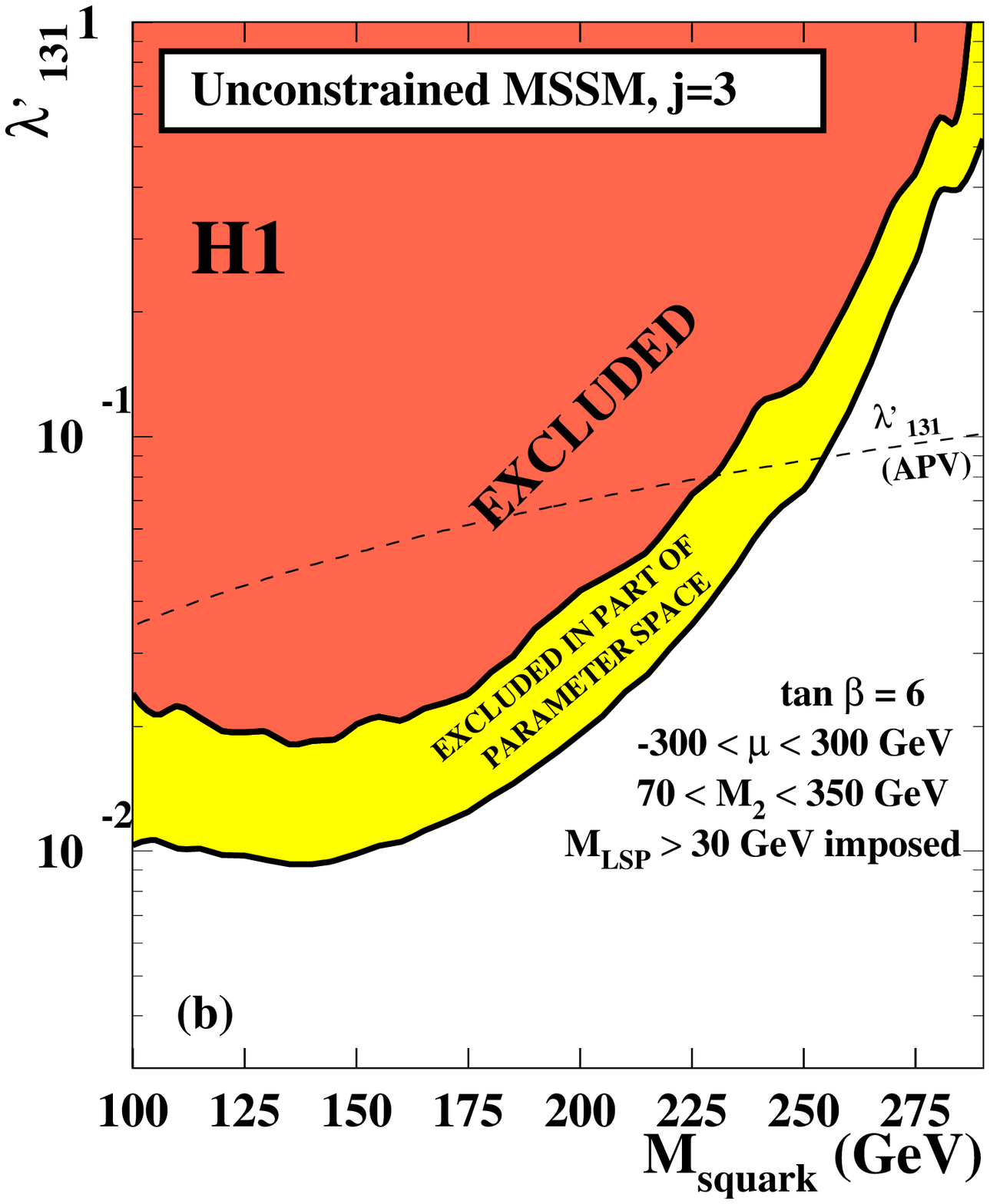}
\includegraphics*[width=5.9cm]{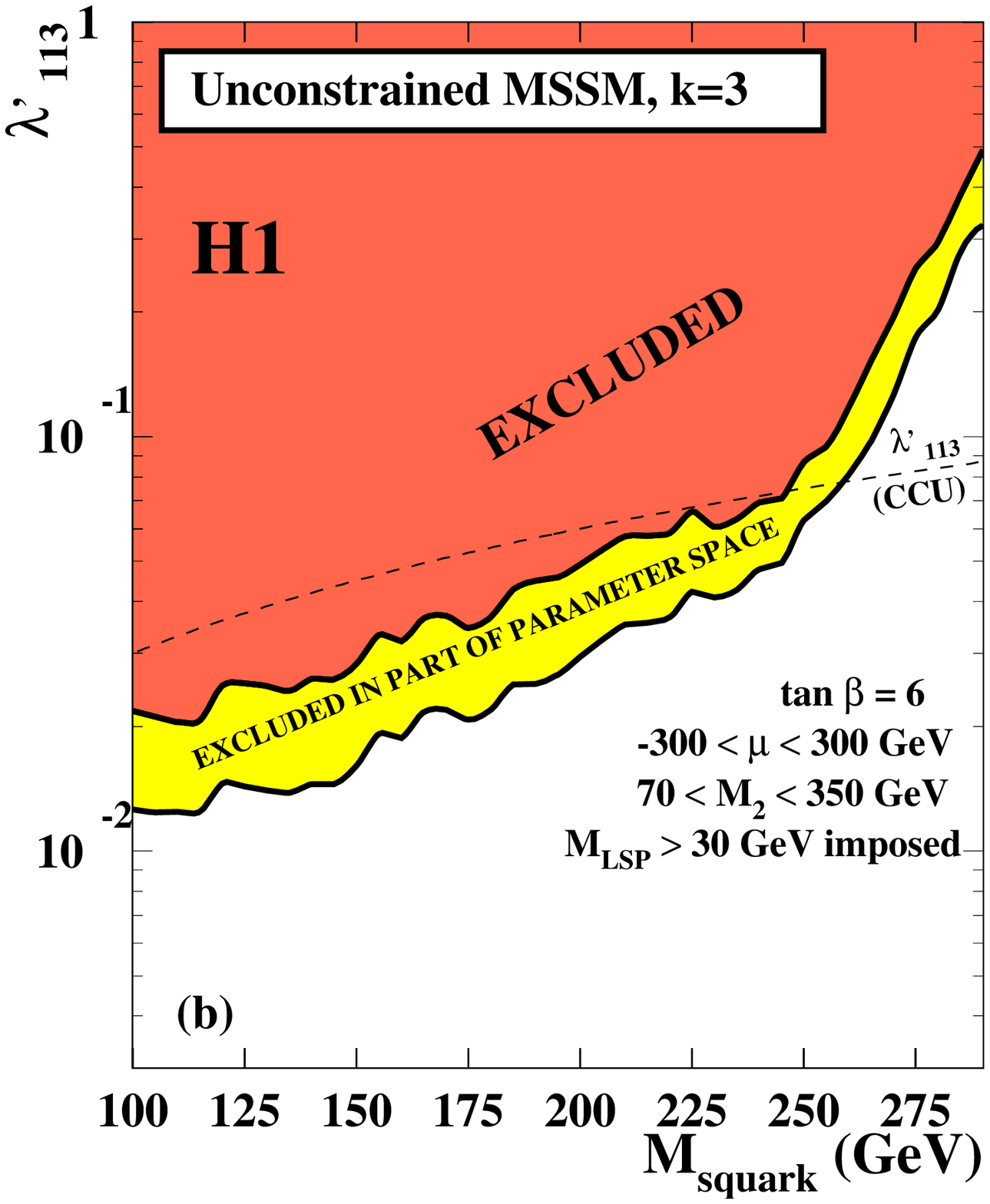}
\caption{Constraint on \Rp\ violating SUSY derived by H1: Exclusion limits ($95\,\%$ CL) on (left) $\lambda'_{131}$, i.e. stop production, and (right) $\lambda'_{113}$, i.e. sbottom production, as a function of the squark mass. The strongest and the weakest limit on $\lambda'$ as derived from a scan of the SUSY parameter space are shown.}
\label{fig:sqlim}
\end{center}
\end{figure} 

\begin{figure}[t] 
\begin{center}
\includegraphics*[width=6.2cm]{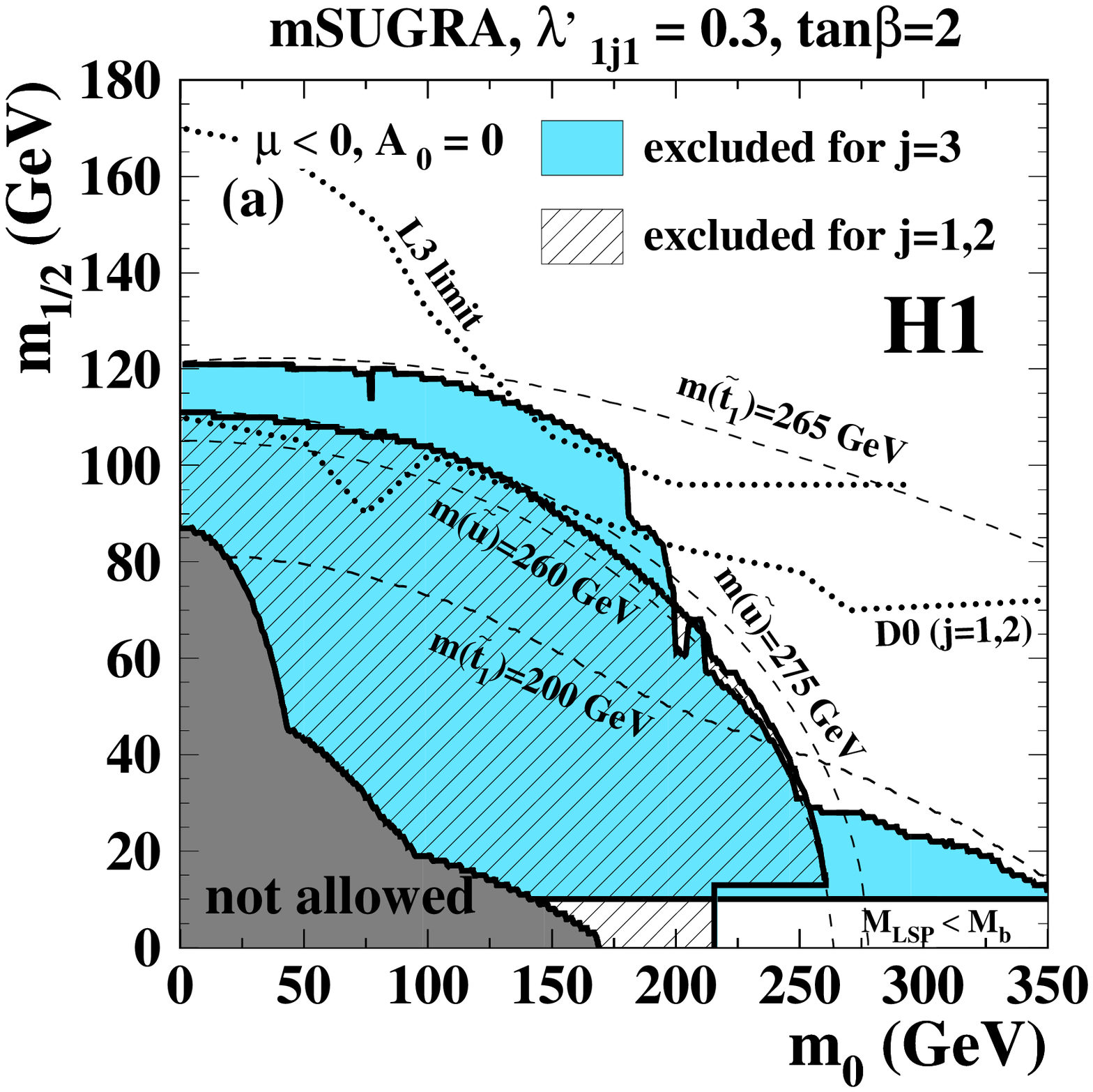}
\includegraphics*[width=6.2cm]{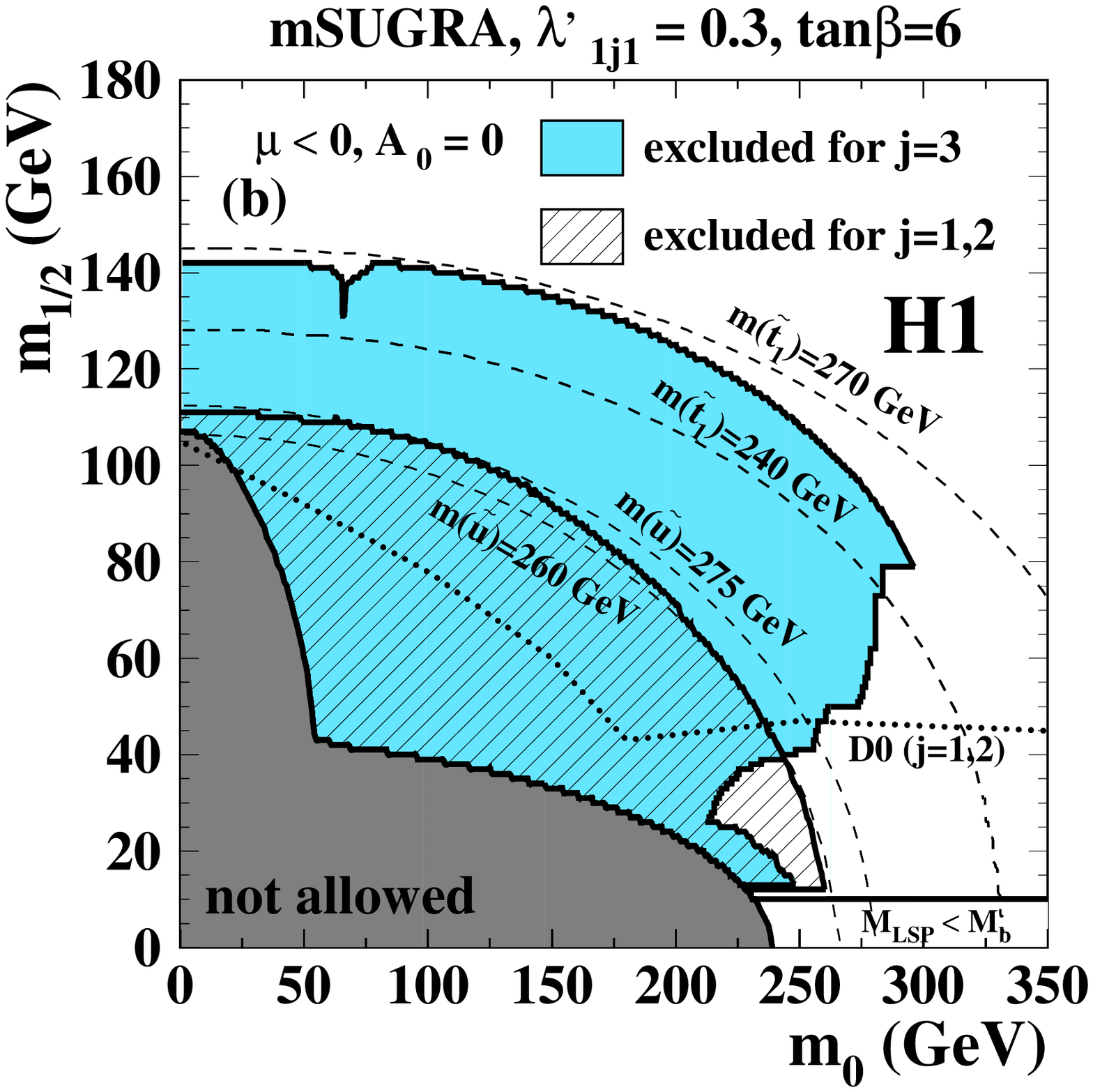}
\caption{
Constraints ($95\,\%$ CL) on minimal Supergravity with non-vanishing $\lambda'$: Exclusion area in the $(m_0,m_{1/2})$-plane for fixed values of $\tan\beta$ and $\lambda'$.}
\label{fig:sugra}
\end{center}
\end{figure}

Exclusion limits on the Yukawa coupling $\lambda'$ as a function of the squark mass have been derived. 
Figure~\ref{fig:sqlim} shows the results for the example couplings $\lambda'_{131}$, corresponding to stop production, and $\lambda'_{113}$, corresponding to sbottom production, interpreted in a MSSM inspired SUSY model, where it is assumed that the gaugino masses depend on the MSSM soft terms, but the sfermion masses are free.
The analysis of the many channels with similar selection efficiencies allows to set limits with only little dependence on the MSSM parameters. 
This is demonstrated by the strongest and the weakest limit on $\lambda'$, indicated in figure~\ref{fig:sqlim} by the red and yellow region, obtained from a parameter scan.
For a Yukawa coupling of electromagnetic strength ($\lambda'=0.3$) squark masses up to $275$\,GeV are excluded.
At lower squark masses the indirect exclusion bounds from low energy experiments are improved by the H1 results.

In addition the results are interpreted in the framework of minimal Supergravity (mSUGRA)~\cite{msugra} with non-vanishing $\lambda'$.
In this model common masses for sfermions and gauginos, $m_0$ and $m_{1/2}$, are assumed at the GUT scale.
Here the electroweak symmetry breaking is driven by radiative corrections.
Figure~\ref{fig:sugra} shows the excluded area in the ($m_{0},m_{1/2}$)--plane for fixed values of $\tan\beta$ and $\lambda'_{1j1}$.
The $\lambda'$ dependent H1 limit follows roughly the squark mass isocurve excluding the region with squark masses below 270\,GeV for $\lambda'=0.3$.
The dependence of the results on $\tan\beta$ was found to be low in the region $\tan\beta<50$. 
The limits from LEP and the Tevatron in this model (mSUGRA + $\lambda'\ne 0$) are of the same order of magnitude, but independent of $\lambda'$. 

The limiting factor of searches for resonantly produced new particles at HERA, like leptoquarks and squarks in \Rp\ SUSY, is the centre-of-mass energy available in the $ep$ collisions. 
For this reason a possible increase of the beam energies would lead to significantly better limits, whereas the improvements of the limits due to higher integrated luminosities are rather small.

\subsection{Bosonic Decays of Stops}

The observation of events with isolated leptons and missing transverse momentum (cf. section~3) triggered a number of new ideas for searches beyond the SM.
One possible explanation for the observed excess~\cite{stoptheo} is the bosonic decay of the stop, the supersymmetric partner of the top quark.
The stop could be produced resonantly in $e^+p$ collisions through an \Rp\ violating reaction via the coupling $\lambda'_{131}$ as discussed above.
The following bosonic decay of the stop into a $W^+$ and a sbottom, which subsequently decays in an \Rp\ reaction to $\bar{\nu_e}+d$, could lead to the observed event topology if the $W^+$ decays leptonically.
To open this decay chain two conditions must be fulfilled. First, decays of the stop to gauginos must be kinematically forbidden (high gaugino masses).
Second, the sbottom quark must be significantly lighter than the stop to allow its decay to $\tilde{b}+W^+$.

\begin{figure}[t] 
\begin{center}
\includegraphics*[width=7.5cm]{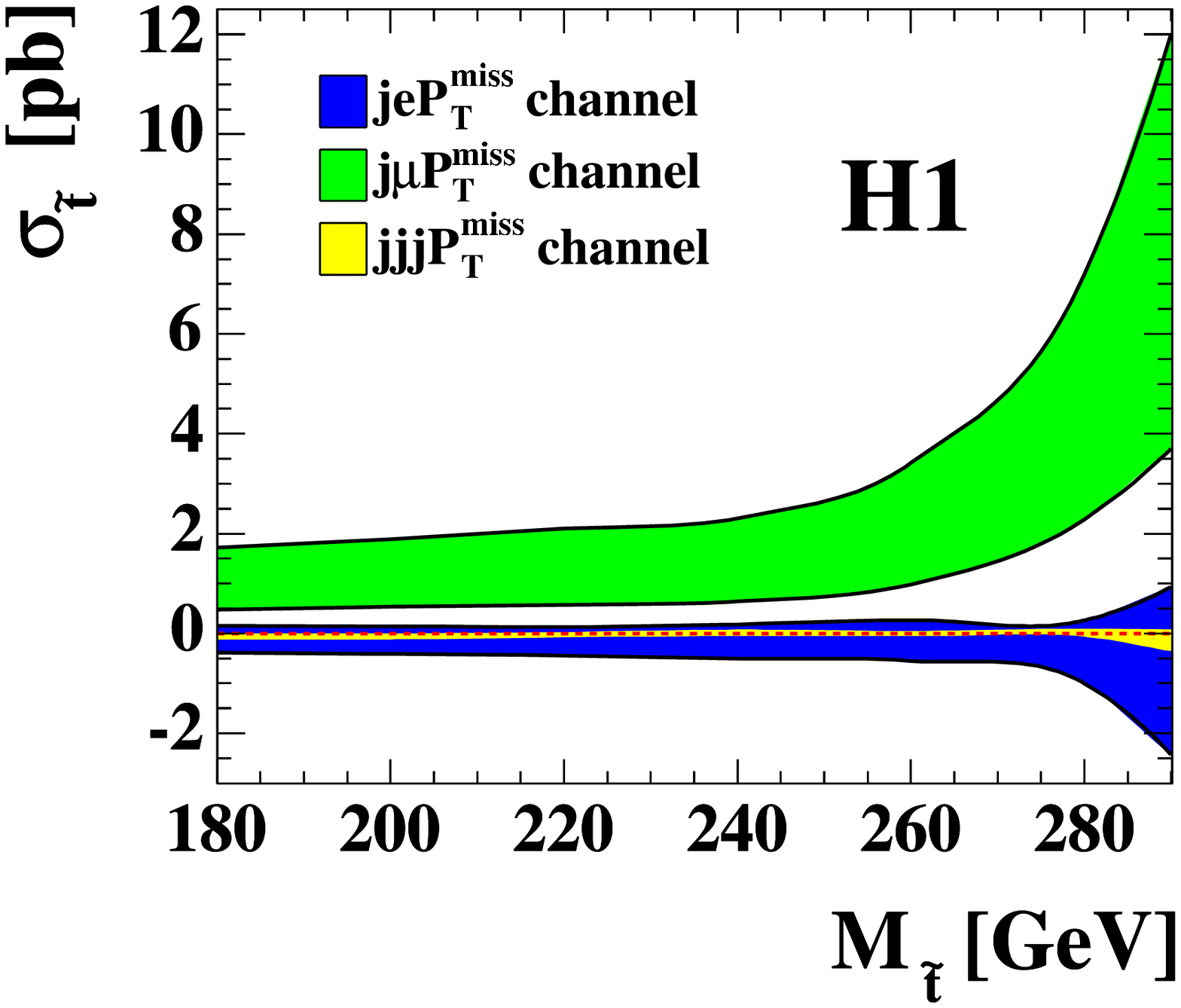}
\includegraphics*[width=6.3cm]{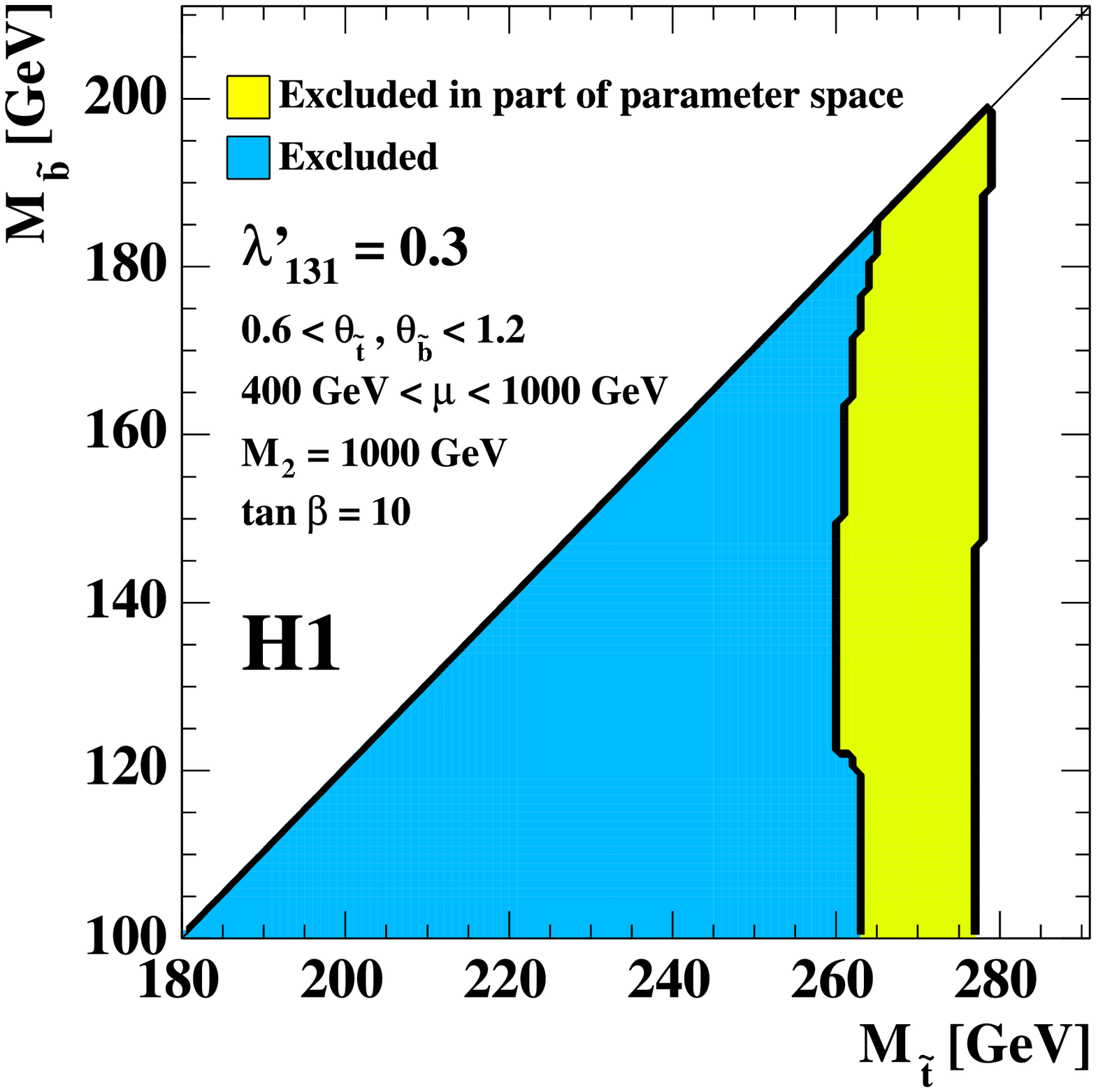}
\caption{
(left) Allowed signal cross section with the corresponding error band for final states of bosonic stop decays in \Rp\ SUSY with non-vanishing $\lambda'$ as measured by H1; (right) Exclusion area in the ($m_{\rm stop},m_{\rm sbottom}$)-plane for a fixed value of $\lambda'_{131}$.}
\label{fig:stop}
\end{center}
\end{figure}

The bosonic decay of the stop leads to the following topologies: $jep_{T,\rm miss}$, $j\mu p_{T,\rm miss}$ and $jjjp_{T,\rm miss}$ where $j$ denotes hadronic jets.
In addition the stop can decay violating $R_p$ directly via $\lambda'_{131}$ into $eq$.
These topologies have been investigated by the H1 experiment in a dedicated analysis~\cite{h1stop}.
All topologies are found to be in agreement with the SM prediction, except the $j\mu p_{T,\rm miss}$ channel, where a clear excess is observed, corresponding to the known excess discussed in section~3.
For all channels the allowed stop production cross section $\sigma_{\tilde{t}}$ was calculated taking into account the number of expected and observed events as a function of the invariant mass of the decay products.
Figure~\ref{fig:stop}~(left) illustrates these cross sections with the corresponding uncertainties.
It can clearly be seen that the excess in the $j\mu p_{T,\rm miss}$ channel is not supported by the other channels whose allowed cross section is consistent with a non-observation, i.e. $\sigma_{\tilde{t}}=0$.
This leads to the conclusion that the isolated lepton events can not be interpreted as a result of bosonic stop decays in \Rp\ violating SUSY.

Therefore exclusion limits have been determined. 
In figure~\ref{fig:stop}~(right) the excluded area in the plane spanned by the stop and sbottom masses is shown for a fixed value of the coupling $\lambda'_{131}$.
Again the relevant SUSY parameters, namely the mixing angles in the stop and sbottom sectors and the ``mass'' term $\mu$, have been scanned.
Stop masses up to about 270\,GeV are excluded for $\lambda'=0.3$ almost independently of the sbottom mass.

\subsection{Neutralino Production via $t$-Channel Slepton Exchange}
For the first time at HERA the H1 Collaboration has searched for new physics in events containing an isolated photon, a jet and missing transverse energy~\cite{h1grav}.
The data are found to be in agreement with the SM where the main contribution comes from radiative CC DIS with a photon and a neutrino in the final state.
The event topology can be interpreted in \Rp\ SUSY ($\lambda'\ne 0$).
A neutralino could be produced via \Rp\ $t$-channel exchange of sleptons in $e^{\pm}p$ collisions.
This production mode gives sensitivity to SUSY models with high squark masses, like models with gauge-mediated SUSY breaking.
In such models the gravitino is the LSP and the neutralino (NLSP) decays into a photon and a light gravitino.
Hence the analysed final state could be observed.
Information about $\lambda'$ can be derived independent of the squark sector.
Exclusion bounds of up to about 110\,\GeV\ on the neutralino mass and 160\,\GeV\ on the mass of the supersymmetric partner of the left-handed electron are set for $\lambda'_{1jk}=1$.

\section{First Results from HERA\,II}
In 2000/2001 the HERA collider was upgraded to deliver higher luminosities and longitudinally polarised electron beams for the two interaction regions.
The first data taking periods in 2002 and 2003 suffered from large background rates, resulting in a limited accumulated data sample.
These starting problems are solved and HERA now delivers about 10\,$\rm pb^{-1}$ per month giving hope to collect 700\,$\rm pb^{-1}$ by mid 2007.
The new data from HERA\,II are eagerly awaited to clarify the situation in the area of the outstanding event signatures observed in the HERA\,I dataset as discussed in section 3.

So far an integrated luminosity of about 50\,$\rm pb^{-1}$ of $e^+p$ collisions has been accumulated. 
H1 already presented first preliminary results based on this dataset~\cite{h1hera2}.
A search for events containing isolated leptons (electrons and muons) and missing transverse momentum is performed using the same selection as for HERA\,I.
The data are found to be consistent with the SM for low values of $p_T^{\rm had}$.
However, for $p_T^{\rm had}>25\,\GeV$ three events in the electron channel and no event in the muon channel are found compared to SM predictions of $0.7\pm0.2$ and $0.8\pm0.1$, respectively.
Thus, this striking event signature is also observed in the new data similar to the previous observation at HERA\,I (cf. table~\ref{tab:iltotnum}).
The probability for the SM expectation to fluctuate to the observed number of events or more in the combined dataset (HERA\,I + II, $163\,{\rm pb}^{-1}$) combining the results in the electron and the muon channel is 0.0022 for $p_T^{\rm had}>25\,\GeV$ compared to 0.0015 reported for the same region in the HERA\,I data alone.

Also the event topology of multiple leptons has been investigated in the new dataset.
No new event with two or three electrons has been found in the new dataset so far.
But the analysis has been extended to also include final states with mixed lepton flavours such as $e\mu$ and $e\mu\mu$ topologies. 
In the full dataset (HERA\,I+II) no $e\mu$ event was found  with a di-lepton mass $M_{e\mu}>100\,\GeV$.
Two $e\mu\mu$ events at high masses were found, one of which has $M_{e\mu}>100\,\GeV$ where only $0.04\pm0.01$ events are expected from the SM.
The other event has $M_{\mu\mu}>100\,\GeV$ with only $0.02\pm0.01$ SM expectation.

More data from HERA\,II are needed to clarify the situation of the outstanding event signatures with leptons observed in the HERA\,I data.
The successful data taking in $e^+p$ collision mode was stopped recently to prepare for $e^-p$ collisions.
The operation of HERA is planned to be resumed in October 2004 with $e^-p$ collisions.

\section{Conclusion}
The HERA collaborations, H1 and ZEUS, have contributed to the search for physics beyond the SM during the last decade.
They have searched for deviations from the SM in the HERA\,I dataset of $e^{\pm}p$ collisions at centre-of-mass energies of up to 319\,GeV. 
In general the data are in good agreement with the expectation.
The search results were used to set constraints in a large variety of models describing new physics.
These constraints are complementary to those obtained at other colliders or by low energy experiments and they often exclude regions not ruled out before.
However, deviations from the SM expectation have been observed for some event topologies, whose interpretation is unclear.
More data from HERA\,II are needed to clarify the situation.
After an upgrade of the collider and its detectors the data taking of HERA\,II has started and the first preliminary results from a small sample of $e^+p$ collisions have been presented.
At the moment the collider prepares for $e^-p$ collisions.
With the current machine performance 700\,$\rm pb^{-1}$ of data are expected for each experiment by mid 2007, thereby offering a better discovery potential and the possibility to improve the current exclusion limits significantly in some models of new physics.

\bibliographystyle{plain}

\begin{thebibliography}{99}
%






\bibitem{h1nccc}C.\,Adloff{\it\,et\,al.}\,[H1 Collab.], {\it Eur. Phys. J. }{\bf C\,30}\,(2003)\,1, [hep-ex/0304003];\\
C.\,Adloff{\it\,et\,al.}\,[H1 Collab.], {\it Eur. Phys. J. }{\bf C\,19}\,(2001)\,269, [hep-ex/0012052];\\
C.\,Adloff{\it\,et\,al.}\,[H1 Collab.], {\it Eur. Phys. J. }{\bf C\,13}\,(2000)\,609, [hep-ex/9908059]. 
\bibitem{znccc} S.\,Chekanov{\it\,et\,al.}\,[ZEUS Collab.], ``High $Q^2$ neutral current cross section in $e^+p$ deep inelastic scattering at $\sqrt{s}=318\,\GeV$'', accepted by {\it Phys. Rev. }{\bf D}, [hep-ex/0401003];  
S.\,Chekanov{\it\,et\,al.}\,[ZEUS Collab.], {\it Eur. Phys. J. }{\bf C\,32}\,(2003)\,1, [hep-ex/0307043];
S.\,Chekanov{\it\,et\,al.}\,[ZEUS Collab.], {\it Phys. Lett. }{\bf B\,539}\,(2002)\,197, [hep-ex/0205091], and erratum ibid. B\,552(2003)\,308;
S.\,Chekanov{\it\,et\,al.}\,[ZEUS Collab.], {\it Eur. Phys. J. }{\bf C\,28}\,(2003)\,175, [hep-ex/0208040];
S.\,Chekanov{\it\,et\,al.}\,[ZEUS Collab.], {\it Eur. Phys. J. }{\bf C\,21}\,(2001)\,443, [hep-ex/0105090];
J.\,Breitweg{\it\,et\,al.}\,[ZEUS Collab.], {\it Eur. Phys. J. }{\bf C\,12}\,(2000)\,411, [hep-ex/9907010] and erratum ibid. {\bf C\,27}\,(2003)\,305.

\bibitem{cilang} E.\,J.\,Eichen, K.\,D.\,Lane, M.\,E.\,Peskin, {\it Phys. Rev. Lett.} {\bf 50}\,(1983)\,811;\\
R.\,R\"uckl, {\it Phys. Lett.} {\bf B\,129}\,(1983)\,363 and {\it Nucl. Phys.} {\bf B\,234}\,(1984)\,91.
\bibitem{h1ci} C.\,Adloff{\it\,et\,al.}\,[H1 Collab.], {\it Phys. Lett.} {\bf B\,568}\,(2003)\,35, [hep-ex/0305015].
\bibitem{zci} S.\,Chekanov{\it\,et\,al.}\,[ZEUS Collab.], {\it Phys. Lett.} {\bf B.\,591}\,(2004)\,23, [hep-ex/0401009].  
\bibitem{arkani} N.\,Arkani-Hamed, S.\,Dimopoulos, G.\,Dvali, {\it Phys. Lett.} {\bf B\,429}\,(1998)\,263.
\bibitem{cdfci} F.\,Abe{\it\,et\,al.}\,[CDF Collab.], {\it Phys. Rev. Lett.} {\bf 79}\,(1997)\,2198.
\bibitem{l3ci} M.\,Acciarri{\it\,et\,al.}\,[L3 Collab.], {\it Phys. Lett.} {\bf B\,489}\,(2000)\,81.

\bibitem{h1gen} A.\,Aktas{\it\,et\,al.}\,[H1 Collab.], ``A general search for new phenomena in ep scattering at HERA'', submitted to {\it Phys. Lett.} {\bf B}, [hep-ex/0308044].
%
\bibitem{h1isolep98} C.\,Adloff{\it\,et\,al.}\,[H1 Collab.], {\it Eur. Phys. J.} {\bf C\,5}\,(1998)\,575, [hep-ex/9806009].
\bibitem{h1isolep} C.\,Adloff{\it\,et\,al.}\,[H1 Collab.], {\it Phys. Lett.} {\bf B\,561}\,(2003)\,241, [hep-ex/0301030].
\bibitem{h1mulel} A.\,Aktas{\it\,et\,al.}\,[H1 Collab.], {\it Eur. Phys. J.} {\bf C\,31}\,(2003)\,17, [hep-ex/0307015].
\bibitem{zmullep} ZEUS Collab., ICHEP 2002 contributed paper, abstract 910.
\bibitem{wprod} K.\,P.\,Diener{\it\,et\,al.}, {\it Eur. Phys. J.} {\bf C\,25}\,(2002)\,405, [hep-ph/0203269]. 
\bibitem{ztop} S.\,Chekanov{\it\,et\,al.}\,[ZEUS Collab.], {\it Phys. Lett.} {\bf B\,599}\,(2003)\,153, [hep-ex/0302010], and addendum in DESY-03-188.
\bibitem{ztau} S.\,Chekanov{\it\,et\,al.}\,[ZEUS Collab.], {\it Phys. Lett.} {\bf B\,583}\,(2004)\,41, [hep-ex/0311028].
\bibitem{h1tau} H1 Collab., ICHEP 2004 contributed paper, abstract 12-0188.
\bibitem{toptheo} V.\,F.\,Obraztsov, S.\,Slabospitsky, O.\,Yushchenko, {\it Phys. Lett.} {\bf B\,426}\,(1998)\,393.
\bibitem{h1top} A.\,Aktas{\it\,et\,al.}\,[H1 Collab.], {\it Eur. Phys. J.} {\bf C\,33}\,(2004)\,9, [hep-ex/0310032].
\bibitem{h1mulmu} A.\,Aktas{\it\,et\,al.}\,[H1 Collab.], {\it Phys. Lett.} {\bf B\,583}\,(2004)\,28, [hep-ex/0311015].
\bibitem{lrs} R.\,N.\,Mohapatra and R.\,E.\,Marshak, {\it Phys. Rev. Lett.} {\bf C\,44}\,(1980)\,1316.
\bibitem{h1dch} H1 Collab., ICHEP 2004 contributed paper, abstract 12-0767.


\bibitem{brw} W.\,Buchm\"uller, R.\,R\"uckl, D.\,Wyler, {\it Phys. Lett.} {\bf B\,191}\,(1987)\,442.
\bibitem{zlq} S.\,Chekanov{\it\,et\,al.}\,[ZEUS Col.], {\it Phys. Rev.} {\bf D\,68}\,(2003)\,052004, [hep-ex/0304008].
\bibitem{h1lq} H1 Collab., EPS 2003 contributed paper, abstract 105.
\bibitem{h1lfv} H1 Collab., ICHEP 2004 contributed paper, abstract 12-0766.
\bibitem{zlfv} S.\,Chekanov{\it\,et\,al.}\,[ZEUS Col.], {\it Phys. Rev.} {\bf D\,65}\,(2002)\,092004, [hep-ex/0201003].  

\bibitem{h1rpc} S.\,Aid{\it\,et\,al.}\,[H1 Collab.], {\it Phys. Lett.} {\bf B\,380}\,(1996)\,461, [hep-ex/9605002].
\bibitem{zrpc} S.\,Chekanov{\it\,et\,al.}\,[ZEUS Collab.], {\it Phys. Lett.} {\bf B\,434}\,(1998)\,214, [hep-ex/9806019].
\bibitem{dreiner} J.\,Butterworth, H.\,Dreiner, {\it Nucl. Phys.} {\bf B\,397}\,(1993)\,3, [hep-ph/9211204].
\bibitem{h1rpv} A.\,Aktas{\it\,et\,al.}\,[H1 Collab.], {\it Eur. Phys. J.} {\bf C\,36}\,(2004)\,425, [hep-ex/0403027].
\bibitem{msugra} G.\,L.\,Kane{\it\,et\,al.}, {\it Phys. Rev.} {\bf D\,49}\,(1994)\,6173, [hep-ph/9312272].

\bibitem{stoptheo} T.\,Kon{\it\,et\,al.}, {\it Mod. Phys. Lett.} {\bf A\,12}\,(1997)\,3143, [hep-ph/9707355].
\bibitem{h1stop} A.\,Aktas{\it\,et\,al.}\,[H1 Collab.], ``Search for bosonic stop decays in R-Parity violating supersymmetry in $e^+p$ collisions at HERA'', acc. by {\it Phys. Lett.} {\bf B}, [hep-ex/0405070].

\bibitem{h1grav} A.\,Aktas{\it\,et\,al.}\,[H1 Collab.], ``Search for Light Gravitinos in Events with Photons and Missing Transverse Momentum at HERA'', to be subm. to {\it Phys. Lett.} {\bf B}.


\bibitem{h1hera2} H1 Collab., ICHEP 2004 contributed paper, abstract 12-0765.
%
\end{thebibliography}

\noindent  

\end{document}